\documentclass[lettersize,journal]{IEEEtran}
\IEEEoverridecommandlockouts
\usepackage{makecell}
\usepackage{array}
\usepackage{graphicx,amssymb,amsmath}
\usepackage{multicol}
\usepackage[noadjust]{cite}
\usepackage{setspace}
\usepackage{subfigure}
\usepackage{graphicx}
\usepackage{float}
\usepackage{url}
\usepackage{stfloats}
\usepackage{amsthm,pifont}
\usepackage{flushend}
\usepackage{cases,subeqnarray}
\usepackage{bm,multirow,bigstrut}
\usepackage{amsmath, amsthm, amssymb}
\usepackage{textcomp}
\usepackage{latexsym,bm}
\usepackage{booktabs}
\usepackage{xcolor}
\usepackage{mathtools}
\usepackage{dsfont}
\usepackage{extarrows}
\usepackage{epsfig}
\usepackage{epstopdf}
\usepackage[noend]{algpseudocode}
\usepackage{algorithmicx,algorithm}
\renewcommand{\algorithmicrequire}{\textbf{Input:}}
\renewcommand{\algorithmicensure}{\textbf{Output:}}

\usepackage{cite}
\usepackage{bm}
\usepackage{cleveref}
\usepackage{multicol}       
\usepackage{multirow}       
\usepackage{array}          
\usepackage{colortbl}
\usepackage{makecell}
\definecolor{crimson}{RGB}{192,0,0}         
\definecolor{navy}{RGB}{47,85,151}         
\usepackage{bbding}
\usepackage{graphicx}
\usepackage{booktabs}
\usepackage{algorithm}
\usepackage{algpseudocode}

\theoremstyle{plain}

\theoremstyle{plain}
\newtheorem{rem}{Remark}

\usepackage{amsmath}

\IEEEoverridecommandlockouts
\begin{document}

\title{Joint Precoding and AP Selection for Energy Efficient RIS-aided Cell-Free Massive MIMO Using Multi-agent Reinforcement Learning}
\author{Enyu Shi, Jiayi~Zhang,~\IEEEmembership{Senior Member,~IEEE}, Ziheng Liu, Yiyang Zhu, Chau Yuen,~\IEEEmembership{Fellow,~IEEE}, Derrick Wing Kwan Ng,~\IEEEmembership{Fellow,~IEEE}, Marco Di Renzo,~\IEEEmembership{Fellow,~IEEE}, and Bo Ai,~\IEEEmembership{Fellow,~IEEE}

\thanks{E. Shi, J. Zhang, Z. Liu, Y. Zhu, and B. Ai are with the School of Electronics and Information Engineering, Beijing Jiaotong University, Beijing 100044, P. R. China. (e-mail: jiayizhang@bjtu.edu.cn).}
\thanks{C. Yuen is with the School of Electrical and Electronics Engineering, Nanyang Technological University, Singapore 639798, Singapore (e-mail: chau.yuen@ntu.edu.sg).}
\thanks{D. W. K. Ng is with the School of Electrical Engineering and Telecommunications, University of New South Wales, NSW 2052, Australia. (e-mail: w.k.ng@unsw.edu.au).}

\thanks{M. Di Renzo is with Universit\'{e} Paris-Saclay, CNRS, CentraleSup\'{e}lec, Laboratoire des Signaux et Syst\`{e}mes, 3 Rue Joliot-Curie, 91192 Gif-sur Yvette, France. (e-mail: marco.di-renzo@universite-paris-saclay.fr).}

}

\maketitle
\begin{abstract}
Cell-free (CF) massive multiple-input multiple-output (mMIMO) and reconfigurable intelligent surface (RIS) are two advanced transceiver technologies for realizing future sixth-generation (6G) networks.
In this paper, we investigate the joint precoding and access point (AP) selection for energy efficient RIS-aided CF mMIMO system. To address the associated computational complexity and communication power consumption, we advocate for user-centric dynamic networks in which each user is served by a subset of APs rather than by all of them. Based on the user-centric network, we formulate a joint precoding and AP selection problem to maximize the energy efficiency (EE) of the considered system. To solve this complex nonconvex problem, we propose an innovative double-layer multi-agent reinforcement learning (MARL)-based scheme.
Moreover, we propose an adaptive power threshold-based AP selection scheme to further enhance the EE of the considered system. To reduce the computational complexity of the RIS-aided CF mMIMO system, we introduce a fuzzy logic (FL) strategy into the MARL scheme to accelerate convergence. The simulation results show that the proposed FL-based MARL cooperative architecture effectively improves EE performance, offering a 85\% enhancement over the zero-forcing (ZF) method, and achieves faster convergence speed compared with MARL. It is important to note that increasing the transmission power of the APs or the number of RIS elements can effectively enhance the spectral efficiency (SE) performance, which also leads to an increase in power consumption, resulting in a non-trivial trade-off between the quality of service and EE performance.
\end{abstract}


\IEEEpeerreviewmaketitle

\section{Introduction}
\IEEEPARstart{W}{ith} the relentless growth of wireless data traffic for sixth-generation (6G) networks, the demand for ultra-high data rates and energy efficiency, extremely high reliability and low latency, as well as global coverage and connectivity has become increasingly critical \cite{wang2023road}. To address these requirements, more efficient and reliable advanced transceiver technologies and network paradigms are needed \cite{you2024next}. Over the past few decades, numerous emerging technologies have emerged, such as massive multiple-input multiple-output (mMIMO) \cite{larsson2014massive}, unmanned aerial vehicle (UAV) communications \cite{wu2021comprehensive}, network densification \cite{bjornson2016deploying}, and so on. Among these advancements, mMIMO stands out as particularly enticing, garnering considerable attention because of its capability to ensure outstanding quality of service (QoS) for numerous users in networks. 
Additionally, mMIMO transceiver technologies boast the benefit of low fronthaul demands, given that the base station (BS) antennas are deployed in a compact array configuration. Despite these advancements, conventional cellular networks grapple with substantial inter-cell interference challenges. Specifically, users located at cell borders contend with both elevated inter-cell interference and path loss, undermining the overall performance of the system \cite{buzzi2019user}. Consequently, innovative signal processing methods and network paradigms are imperative to address the inherent inter-cell interference issues associated with conventional cellular network deployments.

\subsection{Related Works}
Cell-free (CF) mMIMO has recently emerged as a groundbreaking advanced transceiver technology poised to shape the landscape of future wireless communication systems for 6G \cite{zhang2020prospective}. In essence, CF mMIMO embodies a network architecture characterized by a plethora of geographically dispersed access points (APs) intricately linked to a central processing unit (CPU) for signal processing. This interconnected system infrastructure serves all user equipments (UEs) through the adept utilization of simultaneous spatial multiplexing across identical time-frequency resources \cite{ngo2017cell}. CF mMIMO can efficiently address the inherent challenges of inter-cell interference, which is the primary performance bottleneck in dense cellular networks. Therefore, extensive investigations into numerous crucial aspects and fundamentals of CF mMIMO have been undertaken to realize the full potential of this architecture. 
For instance, in \cite{nayebi2017precoding}, the precoding and power optimization method was studied to maximize the minimum signal-to-interference-plus-noise ratio (SINR) in CF mMIMO systems. Besides, in \cite{bashar2019energy}, the authors investigated the maximum energy efficiency (EE) with the iterative algorithm. Despite its advantages, CF mMIMO still struggles to guarantee adequate QoS in challenging propagation conditions, such as unfavorable scattering environments or significant signal attenuation resulting from substantial physical obstructions \cite{van2021reconfigurable}. Meanwhile, with the increase in the number of APs and antennas, the associated transceiver energy consumption of CF mMIMO networks surges, posing significant challenges to fulfill the requirements of green communication \cite{verma2020toward}. Therefore, other advanced technologies are required to assist CF systems in satisfying skyrocketing demands for higher capacity, unwavering reliability, and reduced energy consumption.

Recently, reconfigurable intelligent surface (RIS) has emerged as a revolutionary technology with the capability to manipulate radio waves at the electromagnetic level \cite{9140329,10555049,pan2021reconfigurable}. Specifically, the incorporation of numerous reflective and refractive elements in RIS enables effective passive beamforming. By adopting the phase of reflected incoming signals, RIS facilitates the realization of reconfigurable radio propagation environments and wireless channels \cite{zhu2024marl}. Furthermore, RIS is characterized by its low-cost and low-power fabrication, offering high flexibility in deployment to enhance communication quality for users experiencing unsatisfactory channel conditions \cite{cai2021intelligent}. For example, in \cite{zuo2020resource}, the authors investigated the downlink system throughput of RIS-aided non-orthogonal multiple access (NOMA) systems. The simulation results show that RIS can efficiently enhance the NOMA system performance. Besides, the authors in \cite{li2021robust} investigated the utilization of RIS for enabling robust and secure UAV communications. These research works have demonstrated that RIS can be extensively utilized in various scenarios to assist other advanced technologies for improving the performance of next-generation wireless networks.

To further realize the requirements of next-generation communication, numerous studies have focused on harnessing the combined strengths of RIS and CF mMIMO to enhance communication performance \cite{10556753}. In RIS-aided CF mMIMO systems, several critical factors such as the deployment location of AP and RIS, joint precoding design, power control, and other factors play pivotal roles in determining system performance. Among these, the joint precoding design is particularly important. Adopting a reasonable joint precoding design scheme can significantly improve system performance. For example, in \cite{zhang2021joint}, joint AP precoding and RIS beamforming frameworks of the RIS-aided CF mMIMO system were proposed to improve the sum-SE through alternating optimization techniques. Besides, in \cite{ma2022cooperative}, the authors investigated the weighted sum-SE for the considered system with manifold optimization (MO). 
Note that most existing research works investigate the SE as it directly relates to QoS. However, as the power consumption is a major concern, energy efficiency (EE) has also emerged as a critical metric in CF mMIMO systems \cite{bashar2019energy,ozccelikkale2015linear}. For instance, the authors investigated the user association problem to improve the EE performance by leveraging traditional successive convex optimization and deep reinforcement learning (DRL) \cite{ghiasi2022energy}, respectively. Due to its low power consumption characteristics, the introduction of RIS into CF mMIMO systems can further improve EE performance. For example, in \cite{le2021energy}, the alternating optimization of RIS phase shifts and AP precoding was proposed with limited backhaul capacity to maximize EE. Also, a hybrid RIS-aided CF framework was proposed in \cite{lyu2023energy} with a block coordinate descent (BCD)-based algorithm, which can achieve a locally optimal energy-efficient solution. Besides, in \cite{siddiqi2022energy}, the authors decoupled the RIS phase shift problem and power allocation problem by adopting zero-forcing (ZF) beamforming. However, none of these works considered AP selection, which is expected to offer significant enhancements for the EE of the RIS-aided CF mMIMO system \cite{buzzi2019user}. In particular, unlike traditional CF systems, RIS alters the original channel states between APs and UEs, leading to new outcomes in AP selection. Indeed, with the consideration of AP selection, the joint active and passive precoding design of RIS-aided CF mMIMO systems becomes significantly more complex and the decoupling process highly challenging. 

Fortunately, machine learning (ML)/artificial intelligence (AI), as an essential technology, shows its great potential in tackling non-convex optimization problems that are mathematically intractable \cite{ghiasi2022energy,chuang2023deep}. Specifically, the authors in \cite{ghiasi2022energy} investigated the AP selection of CF mMIMO systems to maximize the EE with DRL. Besides, in \cite{huang2020reconfigurable}, the DRL was utilized to facilitate the joint design of transmit beamforming at the BS and the phase shift at the RIS for the cellular network. Furthermore, the authors in \cite{chen2023distributed} proposed a distributed machine learning-based approach for RIS-aided CF mMIMO systems to maximize the sum rate. The above works all adopted ML and DRL to solve complex optimization problems. Recently, multi-agent reinforcement learning (MARL) has been proposed as a novel ML paradigm \cite{zhang2021multi}, which can perfectly fit the characteristics of multiple APs in CF mMIMO systems through treating each AP as an agent processing data. For example, the authors in \cite{rahmani2022multi} and \cite{liu2023double} proposed a MARL-based pilot assignment approach and power control method for CF mMIMO systems, respectively. Besides, in \cite{banerjee2023access}, the authors utilized the MARL to address the AP clustering problem in CF mMIMO systems. Moreover, in \cite{liu2023uplink}, the authors utilized fuzzy logic (FL) combined with MARL to design power control schemes for CF extremely large-scale MIMO systems and pointed out that FL reduces the number of virtual agents through fuzzy actual inputs, thereby reducing the dimensionality and complexity of calculations. Also, in \cite{tilahun2023multi}, a distributed resource allocation in CF mMIMO-enabled mobile edge computing networks was proposed by MARL. Besides, the authors in \cite{abdallah2024multi} proposed the MARL method for beam codebook design in RIS-aided multiple-input single-output (MISO) systems. These research works indicate that the advanced MARL technique can be applied to CF mMIMO systems to effectively address the complex multivariate nonconvex problems while reducing the computational complexity and information exchanges. Based on this, applying the MARL to RIS-aided CF mMIMO systems to guide the complex multivariate resource allocation problem is a promising approach \cite{9733984}.

\subsection{Motivations and Contributions}
Inspired by the aforementioned observations, to fulfill the stringent EE requirement for 6G, this work endeavors to enhance EE by delving into the joint precoding and AP selection within RIS-aided CF mMIMO systems. Confronted with this intricate multi-variable, non-convex resource allocation quandary, the deployment of novel MARL algorithms is proposed as a solution.
More precisely, we propose a novel double-layer MARL framework to jointly optimize the precoding at the APs, the phase shift at the RIS, and the AP selection to maximize the EE of the considered system. Additionally, FL is integrated to streamline the complexity inherent of the MARL algorithm, offering a balanced and efficient strategy for the considered system. Specifically, the contributions of this paper can be summarized as follows.
 
\begin{itemize}
\item We first investigate a practical user-centric RIS-aided CF mMIMO system and formulate an optimization problem for joint precoding and AP selection to maximize the EE. To address this intractable complex multivariate non-convex optimization problem, we develop a centralized training and distributed execution approach based on MARL.

\item Based on the proposed MARL algorithm, we design a two-layer network to address the AP selection and precoding, and RIS beamforming design separately. Specifically, the first layer's AP selection and precoding network is responsible for selecting suitable AP clusters for UEs and designing the precoding matrix. For the AP selection, we propose an adaptive power threshold-based AP selection algorithm. This aims to focus the beam and reduce the power consumption. Then, the second layer network is responsible for RIS beamforming design to further enhance the system performance.

\item To reduce the computational complexity and improve the convergence speed of MARL, we introduce the FL strategy into the MARL algorithm and analyze the complexity of MARL and FL-based MARL algorithms. The results reveal that FL-based algorithms can significantly reduce the computational complexity at the cost of negligible performance degradation. Our results demonstrate the effectiveness of the proposed FL-based MARL approach in terms of enhancing the EE. 
\end{itemize}

The remainder of this paper is structured as follows. In Section \uppercase\expandafter{\romannumeral2}, we first describe the system model of RIS-aided CF mMIMO including the channel model and transmission model. Next, Section \uppercase\expandafter{\romannumeral3} introduces a UE-centric AP selection paradigm and power consumption model. Also, a joint AP selection and precoding for maximizing EE is presented. In Section \uppercase\expandafter{\romannumeral4}, we propose a double-layer architecture, which combines the AP selection and precoding network and the RIS passive beamforming network. Then, numerical results and performance analysis are provided in Section \uppercase\expandafter{\romannumeral5}. Finally, Section \uppercase\expandafter{\romannumeral6} concludes this paper.

\textbf{Notation:} Column vectors and matrices are denoted by boldface lowercase letters $\mathbf{x}$ and boldface uppercase letters $\mathbf{X}$, respectively. The superscripts $\mathbf{x}^{\rm{H}}$, $x^\mathrm{T}$, and $x^\mathrm{*}$ are adopted to represent conjugate transpose, transpose, and conjugate, respectively.
The $\triangleq$, $\left\|  \cdot  \right\|$, and $\left\lfloor  \cdot  \right\rfloor $ denote the definitions, the Euclidean norm, and the truncated argument, respectively. ${\rm{tr}}\left(  \cdot  \right)$, $\mathbb{E}\left\{  \cdot  \right\}$, and ${\rm{Cov}}\left\{  \cdot  \right\}$ denote the trace, expectation and covariance operators, respectively. We use ${\text{diag}}\left( {{a_1}, \cdots ,{a_n}} \right)$ to express a block-diagonal matrix. Also, $\otimes$ and $\odot$ denote the Kronecker product and the element-wise product, respectively. 
Then, $\nabla$ denotes the gradient operation. $\mathbb{B}^n$, $\mathbb{Z}^n$, $\mathbb{R}^n$, and $\mathbb{C}^n$ represent the $n$-dimensional spaces of binary, integer, real, and complex numbers, respectively.
Finally, the $N \times N$ zero matrix and identity matrix are denoted by $\mathbf{0}_{N}$ and $\mathbf{I}_{N}$, respectively.


\section{System Model}\label{se:model}
As illustrated in Fig. 1, we consider a standard time division duplex (TDD)-based RIS-aided CF mMIMO system consisting of $L$ APs, one RIS, and $K$ UEs and denote $\mathcal{L} \triangleq \left\{ {1,2, \ldots ,L} \right\}$ and $\mathcal{K} \triangleq \left\{ {1,2, \ldots ,K} \right\}$. Each AP and UE are equipped with $M$ antennas and a single antenna, respectively. We assume that all the APs serve all the UEs via the same time and frequency resources and are connected to the CPU via fronthaul links. Besides, the RIS is equipped with $N$ reflective elements capable of introducing phase shifts to the incoming signals \cite{yu2021smart} and denote $\mathcal{N} \triangleq \left\{ {1,2, \ldots ,N} \right\}$. The RIS is connected to the CPU for the management of its reflection coefficient vector.

\subsection{Channel Model}
We assume a quasi-static block fading model in which the channels are considered to be frequency-flat and static within each coherence time block. As shown in Fig. 1, the channels are categorized into two types: the direct link from the UE to the AP and the cascaded link through RIS. Specifically, let ${\mathbf{h}}_{lk,d} \in {\mathbb{C}^{M \times 1}}$ denote the direct link channel between AP $l$ and UE $k$. The matrix ${{{\mathbf{G}}_{lr}}} \in {\mathbb{C}^{N \times M}}$ denotes the channel matrix from AP $l$ to the RIS and ${{\mathbf{h}}_{rk}} \in {\mathbb{C}^{N \times 1}}$ denotes the channel from the RIS to UE $k$. Then, the aggregated channel from AP $l$ to UE $k$ is given by 
\begin{align}\label{h_lk}
{\mathbf{h}}_{lk}^{\text{H}} = {\mathbf{h}}_{lk,d}^{\text{H}} +{\mathbf{h}}_{rk}^{\text{H}}{\mathbf{\Theta }}{{\mathbf{G}}_{lr}},
\end{align}
where ${\mathbf{\Theta }} = {\text{diag}}\left( {{e^{j{\theta _1}}},{e^{j{\theta _2}}}, \ldots ,{e^{j{\theta _N}}}} \right) \in {\mathbb{C}^{N \times N}}$ denotes the reflection coefficient vector of the RIS. Without loss of generality, we assume the ideal phase shift model, i.e., ${\theta _n} \in \left[ {0,2\pi } \right), n \in \mathcal{N}$.

\subsection{Transmission Model}
In the considered system, we assume that all the APs transmit the downlink data to all the UEs simultaneously and the downlink precoded symbol ${{\mathbf{x}}_l}\in {\mathbb{C}^{M \times 1}}$ at AP $l$ is written as
\begin{align}\label{x_l}
{{\mathbf{x}}_l} = \sum\limits_{k = 1}^K {{{\mathbf{w}}_{lk}}} {s_k},
\end{align}
where ${s_k} \in \mathbb{C}$ denotes the downlink signal transmitted to UE $k$. We assume that the transmitted symbols have a normalized power, i.e., $\mathbb{E}{ \{{{\left| {{s_k}} \right|}^2}} \} = 1$, and $\mathbb{E}\{ {s_j^ * {s_k}} \} = 0,\,\forall j \ne k$. Also, ${{{\mathbf{w}}_{lk}}} \in {\mathbb{C}^{M \times 1}}$ denotes the precoding vector of UE $k$ at AP $l$.
Then, the received signal ${y_k}$ at UE $k$ can be expressed as 
\begin{align}\label{y_k}
  {y_k} &= \sum\limits_{l = 1}^L {{\mathbf{h}}_{lk}^{\text{H}}{{\text{x}}_l}}  + {z_k} \notag \\
   &= \sum\limits_{l = 1}^L {\sum\limits_{i = 1}^K {\left( {{\mathbf{h}}_{lk,d}^{\text{H}} + {\mathbf{h}}_{rk}^{\text{H}}{\mathbf{\Theta }}_{}^{\text{H}}{{\mathbf{G}}_{lr}}} \right){{\mathbf{w}}_{li}}} {s_i}}  + {z_k}, 
\end{align}
where ${z_k} \sim \mathcal{C}\mathcal{N}\left( {0,{\sigma_{k} ^2}} \right)$ is the additive white Gaussian noise at UE $k$, and $\sigma_{k} ^2$ is the noise power.

\begin{figure*}[t]
\label{Figure 1}
\centering
\includegraphics[scale=0.85]{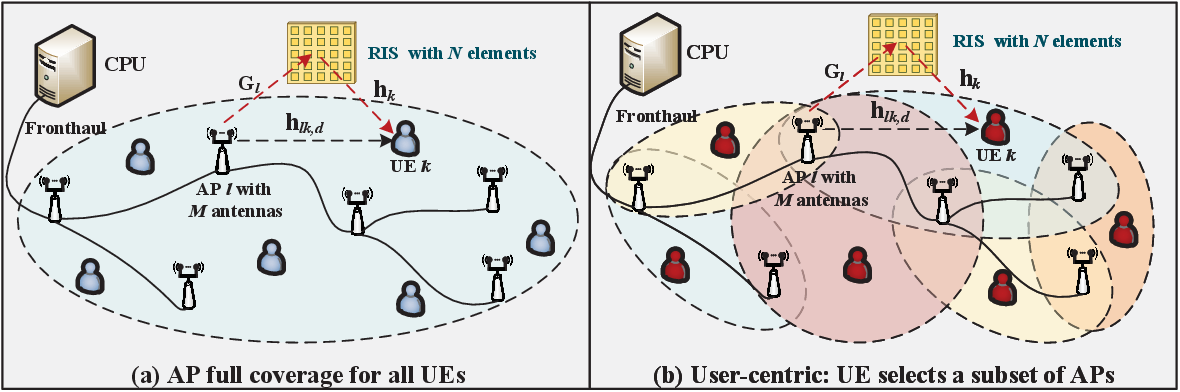}
\caption{Illustration of two representative RIS-aided CF mMIMO systems: (a) AP full coverage for all UEs; (b) User-centric: AP selects a portion of UEs to serve.}
\end{figure*}

\section{Joint AP Selection and Precoding for Energy Efficiency Maximization}
In CF mMIMO systems, some studies assumed that all APs serve all the UEs simultaneously \cite{bjornson2019making,van2021reconfigurable}. However, when the number of APs and UEs is large, this approach will introduce huge costs in information exchanges and will increase the load on the fronthaul link, which goes against the purpose of emerging green communications for next-generation advanced transceivers \cite{bashar2021limited}. In addition, each AP needs to acquire global CSI, which further increases the computational complexity. Hence, we propose a scalable user-centric AP selection paradigm for RIS-aided CF mMIMO systems to reduce the amount of information exchange between the APs and improve the energy efficiency. 

\subsection{UE Centric-based AP Selection Paradigm}
Let ${a_{lk}} \in \left\{ {0,1} \right\}$ denote the binary indicator variable that indicates whether AP $l$ serves UE $k$. Then, the precoded symbol ${{\mathbf{x}}_l}$ in \eqref{x_l} can be formulated as
\begin{align}\label{x_l_a}
{{\mathbf{x}}_l} = \sum\limits_{k = 1}^K {a_{lk}}{{{\mathbf{w}}_{lk}}} {s_k}.
\end{align}
The received signal ${y_k}$ in \eqref{y_k} at UE $k$ can be expressed as 
\begin{align}\label{y_k_a}
  {y_k} 
   \mathop  = \limits^{(a)} \left( {{\mathbf{h}}_{k,d}^{\text{H}} + {\mathbf{h}}_{rk}^{\text{H}}{{\mathbf{\Theta }}^{\text{H}}}{\mathbf{G}}} \right)\sum\limits_{i = 1}^K {{{\mathbf{A}}_i}{{\mathbf{w}}_i}{s_i}}  + {z_k},
\end{align}
where (a) holds by defining ${{\mathbf{h}}_{k,d}} \!=\!\! {[ {{\mathbf{h}}_{1k,d}^{\text{T}},{\mathbf{h}}_{2k,d}^{\text{T}} \ldots ,{\mathbf{h}}_{Lk,d}^{\text{T}}} ]^{\text{T}}} \!\in {\mathbb{C}^{ML \times 1}}$, ${\mathbf{G}} = {\left[ {{{\mathbf{G}}_{r1}},{{\mathbf{G}}_{r2}} \ldots ,{{\mathbf{G}}_{rL}}} \right]^{\text{T}}} \in {\mathbb{C}^{N \times ML}}$, ${{\mathbf{A}}_i} = {\text{diag}}( {\underbrace {{a_{1i}},{a_{1i}} \ldots ,{a_{1i}}}_M, \ldots ,\underbrace {{a_{Li}},{a_{Li}} \ldots ,{a_{Li}}}_M} ) \!\!\in {\mathbb{C}^{ML \times ML}}$, and ${{\mathbf{w}}_i} = {[ {{\mathbf{w}}_{1i}^{\text{T}},{\mathbf{w}}_{2i}^{\text{T}} \ldots ,{\mathbf{w}}_{Li}^{\text{T}}} ]^{\text{T}}} \in {\mathbb{C}^{ML \times 1}}$. Then, the signal-to-interference-plus-noise ratio (SINR) for the transmitted symbol $s_k$ at UE $k$ can be expressed as 
\begin{align}\label{gamma_k}
{\gamma _k} = \frac{{{{\left| {\left( {{\mathbf{h}}_{k,d}^{\text{H}} + {\mathbf{h}}_{rk}^{\text{H}}{\mathbf{\Theta} ^{\text{H}}}{\mathbf{G}}} \right){{\mathbf{A}}_k}{{\mathbf{w}}_k}} \right|}^2}}}{{\sum\nolimits_{i = 1,i \ne k}^K {{{\left| {\left( {{\mathbf{h}}_{k,d}^{\text{H}} + {\mathbf{h}}_{rk}^{\text{H}}{\mathbf{\Theta} ^{\text{H}}}{\mathbf{G}}} \right){{\mathbf{A}}_i}{{\mathbf{w}}_i}} \right|}^2} + {\sigma ^2}} }}.
\end{align}
Thereby, the sum SE of all $K$ UEs is given by
\begin{align}
{\rm{SE}} = \sum\limits_{k = 1}^K {{{\log }_2}\left( {1 + {\gamma _k}} \right)}. 
\end{align}

\subsection{Power Consumption Model and Energy Efficiency}
Without loss of generality, we adopt the power consumption model in \cite{ngo2017total} for CF mMIMO systems and extend it to user-centric RIS-aided CF mMIMO systems. Specifically, the power consumption model includes the following main components: a) the total power of the APs $\left\{ {P_l^{\text{total}}:\forall l} \right\}$; b) the total power of the RIS $\left\{ {P_{\text{RIS}}^{\text{total}}} \right\}$; c) the hardware static power consumption of the UEs $\left\{ {P_k:\forall k} \right\}$ d) the circuit power of the fronthaul links ${P^{\text{fronthaul}}}$ and the static power consumption of the CPU ${P^{\text{CPU}}}$. Mathematically, the total power consumption ${P_{\text{total}}}$ can be expressed as 
\begin{align}\label{P}
{P_{{\text{total}}}} = \sum\limits_{l = 1}^L {P_l^{{\text{total}}}}  + P_{{\text{RIS}}}^{{\text{total}}} + \sum\limits_{k = 1}^K {{P_k}}  +  {P^{{\text{fronthaul}}}} + {P^{\text{CPU}}}.
\end{align}
Then, we model each of these terms in detail. The total power of AP $l \in \mathcal{L}$ is denoted as 
\begin{align}\label{P_l_total}
P_l^{{\text{total}}} = \sum\limits_{k = 1}^K {{{\left( {{a_{lk}}{{\mathbf{w}}_{lk}}} \right)}^{\text{H}}}{a_{lk}}{{\mathbf{w}}_{lk}}}  + P_l^{{\text{static}}},
\end{align}
where the first term is the transmit power and $P_l^{{\text{static}}}$ denotes the hardware static power consumption of AP $l$. 
The dissipated power at an RIS comprising $N$ identical reflecting elements can be denoted as $P_{{\text{RIS}}}^{{\text{total}}} = N{P_{{\text{element}}}}$, where ${P_{{\text{element}}}}$ denotes the power consumption of each RIS element \cite{huang2019reconfigurable}. Note that different from previous cellular systems, e.g., \cite{ngo2017total,van2020power}, in RIS-aided CF mMIMO systems each AP is equipped with a fronthaul link to transmit CSI and precoding coefficients. Therefore, the total power consumption of the fronthaul links is related to the sum SE and the bandwidth, which can be expressed as \cite{ngo2017total}
\begin{align}\label{P_fronthaul}
P_{}^{{\text{fronthaul}}} = \sum\limits_{l = 1}^L {P_l^{{\text{fronthaul}}}},
\end{align}
where $P_l^{{\rm{fronthaul}}} = {P_{0,l}} + {{B}} \cdot {\rm{SE}} \cdot {P_{{\rm{bt}},l}}$ denotes the fronthaul link power consumption of AP $l$. ${P_{0,l}}$ is a fixed power consumption of each fronthaul, ${P_{{\rm{bt}},l}}$ denotes the traffic-dependent power, and $B$ is the system bandwidth. As a result, the total power consumption of the considered RIS-aided CF mMIMO systems can be expressed as
\begin{align}\label{P_total}
{P_{{\text{total}}}} & = \sum\limits_{l = 1}^L {\left( {\sum\limits_{k = 1}^K {{{\left( {{a_{lk}}{{\mathbf{w}}_{lk}}} \right)}^{\text{H}}}{a_{lk}}{{\mathbf{w}}_{lk}}} \! +\! P_l^{{\text{static}}}} \right)} \!+\! N{P_{{\text{element}}}} \notag\\
& + \sum\limits_{k = 1}^K {{P_k}}  + \sum\limits_{l = 1}^L {P_l^{{\text{fronthaul}}}} + {P^{\text{CPU}}}.
\end{align}
Therefore, the system sum SE and EE can be expressed as \eqref{SE} and \eqref{EE} at the top of this page. Note that the EE is determined by the precoding matrices at the AP, the beamforming vector at the RIS, and the AP selection strategy.

\newcounter{mytempeqncnt}
\begin{figure*}[t!]
\normalsize
\setcounter{mytempeqncnt}{1}
\setcounter{equation}{11}
\begin{align}\label{SE}
{\text{SE}}\left( {{\mathbf{w}},{\mathbf{\Theta }},{\mathbf{A}}} \right) = \sum\limits_{k = 1}^K {{{\log }_2}} \left( {1 + \frac{{{{\left| {\left( {{\mathbf{h}}_{k,d}^{\text{H}} + {\mathbf{h}}_{rk}^{\text{H}}{{\mathbf{\Theta }}^{\text{H}}}{\mathbf{G}}} \right){{\mathbf{A}}_k}{{\mathbf{w}}_k}} \right|}^2}}}{{\sum\nolimits_{i = 1,i \ne k}^K {{{\left| {\left( {{\mathbf{h}}_{k,d}^{\text{H}} + {\mathbf{h}}_{rk}^{\text{H}}{{\mathbf{\Theta }}^{\text{H}}}{\mathbf{G}}} \right){{\mathbf{A}}_i}{{\mathbf{w}}_i}} \right|}^2} + {\sigma_{k} ^2}} }}} \right).
\end{align}
\setcounter{equation}{12}
\hrulefill
\end{figure*}
\begin{figure*}[t!]
\normalsize
\setcounter{mytempeqncnt}{2}
\setcounter{equation}{12}
\begin{align}\label{EE}
{\text{EE}}\left( {{\mathbf{w}},{\mathbf{\Theta }},{\mathbf{A}}} \right) &\!=\! \frac{{{{B}} \!\times\! {\text{SE}}\left( {{\mathbf{w}},{\mathbf{\Theta }},{\mathbf{A}}} \right)}}{{{P_{{\text{total}}}}}} \!=\! \frac{{{{B}} \times {\text{SE}}\left( {{\mathbf{w}},{\mathbf{\Theta }},{\mathbf{A}}} \right)}}{{\sum\limits_{l = 1}^L {\left( {\sum\limits_{k = 1}^K {{{\left( {{a_{lk}}{{\mathbf{w}}_{lk}}} \right)}^{\text{H}}}{a_{lk}}{{\mathbf{w}}_{lk}}}  \!+\! P_l^{{\text{static}}}} \right)}  \!+\! N{P_{{\text{element}}}} \!+\! \sum\limits_{k = 1}^K {{P_k}}  \!+\! \sum\limits_{l = 1}^L {P_l^{{\text{fronthaul}}}} \!+\! {P^{\text{CPU}}} }}.
\end{align}
\setcounter{equation}{13}
\hrulefill
\end{figure*}

\subsection{Problem Formulation}
In this paper, our target is to maximize the EE in the considered user-centric RIS-aided CF mMIMO system by jointly designing the precoding matrices at the APs, the phase-shifting vector at the RIS, and the AP-UE selection coefficients. The considered EE maximization problem can be expressed as follows:
\begin{subequations}\label{eq:combined} 
  \begin{align}
  \mathop {\max }\limits_{{\mathbf{w}},{\mathbf{\Theta }},{\mathbf{A}}} \quad &{\text{EE}}\left( {{\mathbf{w}},{\mathbf{\Theta }},{\mathbf{A}}} \right)  \label{eq:first} \\
  \mathrm{s.t.}\quad \;\;&\sum\limits_{k = 1}^K {{{\left\| {{a_{lk}}{{\mathbf{w}}_{lk}}} \right\|}^2}}  \!\!\leqslant \!\!{P_{\text{AP},\max }},\!\!\!\!\quad \forall k \in \mathcal{K},l \in \mathcal{L}, \label{eq:second} \\
  \quad \quad \;\;&{\theta _n} \in \left[ {0,2\pi } \right],\quad \forall n \in \mathcal{N}, \label{eq:third} \\
  \quad \quad \;\;&{a_{lk}} \in \left\{ {0,1} \right\},\quad \forall k \in \mathcal{K},l \in \mathcal{L}, \label{eq:forth}    
  \end{align}\label{Problem}
\end{subequations}
where ${P_{\text{AP},\max }}$ denotes the maximum transmission power of each AP. In detail, constraint \eqref{eq:second} guarantees that the transmit power of the AP remains within the permissible maximum threshold. Also, constraint \eqref{eq:third} accounts for the feasible phase shifts of the reflecting elements. In addition, \eqref{eq:forth} is a binary constraint, indicating the connectivity between each AP and UE. 

The EE optimization problem \eqref{eq:first} is a non-concave mixed integer program, primarily due to the non-convex nature of the objective function \eqref{eq:first}, alongside the sparse constraint \eqref{eq:third} and the binary constraint \eqref{eq:forth}, which renders the problem challenging and complex to solve.
In particular, with the introduction of the AP selection, the problem becomes even more challenging to solve. Hence, there is a need for an innovative, efficient, and low-complexity solution.

\section{Proposed Multi-parallel MARL-based Framework}
\begin{figure*}[t]
\centering
\includegraphics[scale=0.75]{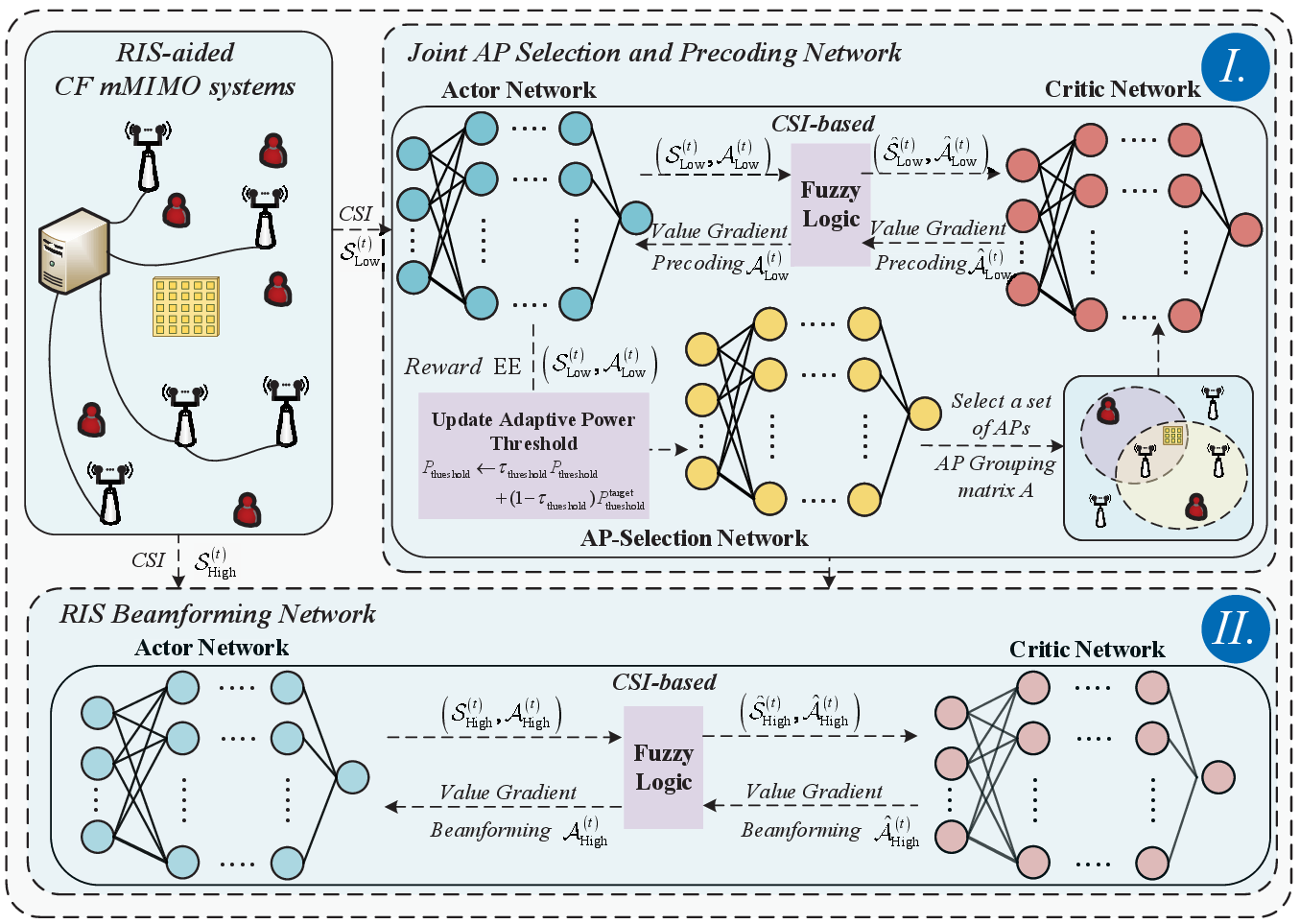}
\caption{Illustration of a double-layer FL-based MARL network, which incorporates the joint AP selection and precoding network and the RIS beamforming network. The first layer network jointly designs the precoding matrix of all APs and adopts an adaptive power threshold-based algorithm for AP selection; The second layer network designs the RIS beamforming based on the output results of the first layer network.}\label{AP_selection_precoding_network}
\end{figure*}

In this section, we propose a double-layer MARL-based framework to address the joint AP selection and precoding design to maximize the system EE of the user-centric RIS-aided CF mMIMO network as shown in Fig.~\ref{AP_selection_precoding_network}. Furthermore, to reduce the computational complexity of high-dimensional networks, we introduce the FL strategy, termed the FL-based multi-agent deep deterministic policy gradient (MADDPG) algorithm. Also, an adaptive power threshold-based AP selection algorithm is proposed to further improve the system performance.

\subsection{FL-based Markov Decision Process Model}
Recently, MARL has gained significant attention for offering effective solutions to resource allocation problems in CF mMIMO systems \cite{banerjee2023access,chuang2023deep}. Most conventional approaches focus on supervised learning, which requires lots of training data and incurs high computational complexity. In contrast, MARL is a disruptive approach that eliminates the need for prior training datasets, positioning it as a viable solution that can adapt to dynamic wireless environments.

In general, MARL is investigated utilizing the Markov Decision Process (MDP), defined by the tuple $<\!\mathcal{S},\mathcal{A},\mathcal{P},\mathcal{R},\gamma\!>$ \cite{chafii2023emergent}, where $\mathcal{S}$ and $\mathcal{A}$ denote the observed state space and the assigned action space of each agent, respectively. The variables $\mathcal{P}:\left( {\mathcal{S},\mathcal{A}} \right) \to \mathcal{S}$ and $\mathcal{R}$ represent the state transition function and the expected reward, respectively. In addition, $\gamma$ denotes the discount factor of the reward. Furthermore, MARL algorithms exhibit three implementation paradigms: value-based, policy-based, and actor-critic approaches. In contrast to value-based and policy-based methodologies, actor-critic methods entail a critic network that accommodates the action-value function within a continuous space. The action network does not necessitate an optimal strategy search on the Q-value table, highlighting the appeal of this architecture for continuous optimization problems, such as MADDPG \cite{liu2023double}. However, due to the presence of numerous distributed APs and RIS with a large number of reflecting elements, the RIS-aided cell-free mMIMO network is vast, resulting in high computational complexity for the traditional MADDPG algorithm. Fortunately, the concept of Federated Learning (FL) was proposed in \cite{hajek2013metamathematics}, and it is verified that by leveraging a reasonable FL algorithm, a significant reduction in computational complexity can be achieved. Building upon this, to reduce the computational complexity of MADDPG, we propose an FL-based MADDPG algorithm that utilizes fuzzy sets to describe the fuzzy relationship between input and output by sacrificing a certain degree of precision and accuracy. In the following, we will introduce the detailed specifics of each layer in the double-layer FL-based MARL network.

\begin{algorithm}[!t]
    \caption{Adaptive Power Threshold-based AP-Selection Algorithm}
    \label{alg:AOA1}
    \renewcommand{\algorithmicrequire}{\textbf{Input:}}
    \renewcommand{\algorithmicensure}{\textbf{Output:}}
    \begin{algorithmic}[1]
    \State \textbf{Initialize} AP agent states $\mathcal{S}_{1}^{(t)},\dots,\mathcal{S}_{L}^{(t)}$ by randomly sampling fuzzy agent states: $\hat{\mathcal{S}}_{1}^{(t)},\dots,\hat{\mathcal{S}}_{F}^{(t)}$, precoding power threshold $P_\text{threshold}$, and power threshold update rate $\tau_\text{threshold}$
    \State Define the target power threshold $P_\text{threshold}^\text{target}$ = 0 and count = 0
    \While {count $\le$ N}
        \State Get AP agent states $\mathcal{S}_{1}^{(t)},\dots,\mathcal{S}_{L}^{(t)}$ with low level observation ${\mathcal{O}}_{AP,l}^{(t)} ={{\mathcal{O}}_{AP,l,1}^{(t)},\dots,{\mathcal{O}}_{AP,l,K}^{(t)}}$ and fuzzy states $\hat{\mathcal{S}}_{1}^{(t)},\dots,\hat{\mathcal{S}}_{F}^{(t)}$ with  $\hat{\mathcal{S}}_{f}^{(t)}=\hat{\mathcal{S}}_{f,1}^{(t)},\dots,\hat{\mathcal{S}}_{f,K}^{(t)}$
        \For {each training step}
            \For {each agent}
                \For {each UE}
                    \State Determine the size of the precoding power value ${\mathcal{S}}_{l,k}^{(t)}$ and the power threshold  $P_\text{threshold}$ to determine whether AP $l$ has selected this UE $k$
                \EndFor
            \EndFor
        \EndFor
        \State Using the overall average to update the target power threshold $P_\text{threshold}^\text{target}$
        \State Update power threshold adopting the target power threshold $P_\text{threshold} \leftarrow \tau_\text{threshold} P_\text{threshold} + (1-\tau_\text{threshold})P_\text{threshold}^\text{target}$
        \State count += 1
    \EndWhile
    \end{algorithmic}
\end{algorithm}

\subsection{Joint AP Selection and Precoding MARL-based Network}
\subsubsection{First-Layer: AP Selection and Precoding Network}
To comprehensively capture the states of the UEs distributed across different locations, we employ an observable method that incorporates partial observation variables and global observation variables, which enables the agent to observe not only the AP-level policy information $O_{L}^{^{(t)}}$, but also the relative positions of all APs, RISs, and UEs at time slot $t$, where $t$ denotes the number of training times in the MARL network. Distributed AP-level control for joint AP selection and precoding, along with a MARL tuple $<\mathcal{S}_\mathrm{AP}^{(t)},\mathcal{A}_\mathrm{AP}^{(t)},r_\mathrm{AP}^{(t)}>$ at slot $t$, can be designed as follows.

\textbf{AP-level agent}: We consider each AP as an agent.

\textbf{AP-level observation space}: It is important to note that AP $l$ only knows its local CSI, while it does not need to be aware of the global CSI of others. Consequently, all observations consist solely of scalar values derived from partial CSI statistics rather than the exact global CSI matrix, significantly reducing the signaling overhead and channel training time. Here, the partial observation for AP $l$ in slot $t$ is denoted as
\begin{align}
    \mathcal{O}_{\mathrm{AP}, l}^{^{(t)}} = \left(\mathbf{A}_{l}^{(t)},\mathbf{h}_{lk,d}^{(t)}, \mathbf{G}_{lr}^{(t)},\mathbf{w}_{l}^{(t)}, {\rm{{EE}}}_{l}^{(t)} \right),
\end{align}
where \textbf{A} is the AP-selection matrix determined by the adaptive power threshold algorithm proposed in Algorithm 1. Specifically, considering that precoding is utilized to improve system performance and eliminate interference, it forms a close relationship with the AP-selection matrix. Therefore, we can set a precoding power threshold and determine the type of UE to choose for each AP based on the precoding matrix output at each AP.

By dividing each dimension of the state space into $F$ unique fuzzy sets, we obtain the AP-level fuzzy state, denoted as $\hat{\mathcal{S}}_{\mathrm{AP}}^{(t)} = ( \hat{\mathcal{S}}_{\mathrm{AP},1}^{(t)},\dots,\hat{\mathcal{S}}_{\mathrm{AP},F}^{(t)} )$, with each ${\hat{\mathcal{S}}_{\mathrm{AP},f}}^{(t)}$ being a random selection from the observed state. For any given $j$-th dimension of the fuzzy set, the corresponding membership function is denoted as $\xi_{\hat{\mathcal{S}}_{\mathrm{AP},f,j}}^{(t)} = {\rm{exp}}(-\frac{1}{d{a} \times n}|\mathcal{S}_{\mathrm{AP}}^{(t)}-\hat{\mathcal{S}}_{\mathrm{AP},f,j}^{(t)}|)$, where $d_{a}$ represents the dimension of the network action space.

\begin{algorithm}[!t]
    \caption{FL-MARL Algorithm for Maximizing EE}
    \label{alg:AOA2}
    \renewcommand{\algorithmicrequire}{\textbf{Input:}}
    \renewcommand{\algorithmicensure}{\textbf{Output:}}
    \begin{algorithmic}[1]
    \State \textbf{Initialize} AP agent states $\mathcal{S}_{1}^{(t)},\dots,\mathcal{S}_{L}^{(t)}$ by randomly sampling fuzzy agent states: $\hat{\mathcal{S}}_{1}^{(t)},\dots,\hat{\mathcal{S}}_{F}^{(t)}$ 
    \State count = 0
    \While {count $\le$ N} 
        \State Evaluate the network actor to decide the downlink precoding and phase shift design: $\hat{\mathcal{A}}_{f}^{(t)} = \pi_{f}\left(\mathcal{S}_{f}^{(t)}\right)$ 
        \State Calculate actual actions $\mathcal{A}_{f}^{(t)} = \sum_{f=1}^{F}\bar\Xi_{f,k}^{(t)} \times \hat{\mathcal{A}}_{f}^{(t)} $
        \State Get actual rewards $r_{i}$ with reward function
        \State Use fuzzy function: $\hat{r}_{f}^{(t)}=\sum_{k=1}^{K} \bar{\Xi}_{f,k}^{(t)} \times r_{k}^{(t)}$ to calculate fuzzy rewards $\hat{r}_{f}$
        \State Update environment
        \State Get next actual states $\mathcal{S}_{i}^{(t)}$ 
        \State Get next fuzzy states $\hat{\mathcal{S}}_{f}^{(t)}$ by fuzz function: $\hat{\mathcal{S}}_{f}^{(t)} = \sum_{k=1}^{K}\bar\Xi_{f,k}^{(t)} \times \hat{\mathcal{S}}_{k}^{(t)}$
        \State Update membership function with $\xi_{\hat{s}_{f,j}}^{(t+1)}$
        \State Store fuzzy experience $<\mathcal{S}^{(t)},\mathcal{A}^{(t)},r^{(t)}>$ in replay buffer $\mathcal{D}_{i}$
        
        \For {each training step}
            \State Randomly sample a mini-batch of $\mathcal{B}_{i}$ transitions uniformly from $\mathcal{D}_{i}$
            \State Calculate the loss function of the joint network $L(\theta)$ with the observed information
            \State Update weights of joint precoding and phase shift critic network
            \State Calculate policy gradient of the two layer actor network $\Delta J(\theta_{Q_{\pi}})$ and update target network
        \EndFor
        \State \textbf{end for}
        \State count += 1
    \EndWhile
    \State \textbf{end while}
    \end{algorithmic}
\end{algorithm}

\addtolength{\topmargin}{0.3cm}
\textbf{AP-level action space}: The action space can be defined as $\mathcal{A}_{\mathrm{AP}} \in \mathcal{L}$. As shown in Fig 2, inspired by traditional alternating optimization, we adopt a joint design policy of AP selection and precoding.

Each fuzzy agent in the AP-level fuzzy system is assigned a policy based on the perceived fuzzy AP-level state $\hat{\mathcal{S}}_{\mathrm{AP}}^{(t)}$. Subsequently, defuzzification is employed to establish a mapping from the fuzzy action $\hat{\mathcal{A}}_{\mathrm{AP}}^{(t)} = ( \hat{\mathcal{A}}_{\mathrm{AP},1}^{(t)}, \dots, \hat{\mathcal{A}}_{\mathrm{AP},F}^{(t)} )$ to the specific action $\mathcal{A}_{\mathrm{AP}}^{(t)}$. Here, we define the relationship between the $p$-th agent and the $f$-th fuzzy agent as $\mathcal{A}_{\mathrm{AP},p}^{(t)}=\sum_{f=1}^{F} \bar{\Xi}{\mathrm{AP},f,p}^{(t)} \times \hat{\mathcal{A}}_{\mathrm{AP},f}^{(t)}$, where $\bar{\Xi}_{\mathrm{AP},f,p}^{(t)}$ represents the normalized mapping relationship, and $\Xi_{\mathrm{AP},f,p}^{(t)} = \prod_{f=j}^{d_{a}} \xi_{\hat{\mathcal{S}}_{\mathrm{AP},f,j}}^{(t)}$.

\begin{rem}
Fuzzy sets and membership functions transform the observed CSI into usable forms within a fuzzy logic system. In MARL, agents perceive fuzzy states $\hat{\mathcal{S}}$ and take fuzzy actions $\hat{\mathcal{A}}$. Membership functions $\xi$ map these fuzzy actions to specific actions by evaluating their relevance in each state, allowing agents to handle uncertainty and imprecision. This enables agents to make informed, flexible decisions, better reflecting real-world scenarios where boundaries are unclear. By leveraging fuzzy sets and membership functions, the system adapts more effectively to dynamic environments.
\end{rem}

\textbf{AP-level reward space}: Subsequent to the reception of the specific action $\mathcal{A}_{\mathrm{AP} }^{(t)}$ by the agents, the corresponding reward $r_{\mathrm{AP}}^{(t)}$ at slot $t$ is ascertained in accordance with the predefined reward function. In order to guarantee equitable service for all UEs, we define the AP-level reward as follows:
\begin{equation}
    r_{\mathrm{AP},l}^{^{(t)}} = \frac{1}{K} \sum_{k \in \mathcal{K}} \mathbb{I}[\sum_{l\in \mathcal{A}_{k}} {\rm{EE}}_{l}^{^{(t)}}], 
\end{equation}
where $\mathbb{I}[x]$ is the indicator function. In the context of interacting with the environment, it is essential to employ fuzzy agents rather than entities. This necessitates the process of fuzzification to derive the fuzzy reward $\hat{r}_{\mathrm{AP}}^{(t)}=( \hat{r}_{\mathrm{AP},1}^{(t)},\dots,\hat{r}_{\mathrm{AP},F}^{(t)} )$ within the framework of reinforcement learning. The fuzzy reward can be expressed as $\hat{r}_{\mathrm{AP},l}^{(t)}=\sum_{k=1}^{K} \bar{\Xi}_{\mathrm{AP},l,k}^{(t)} \times r_{\mathrm{AP},k}^{(t)}$. Therefore, the incorporation of fuzzification is imperative for the successful completion of the reinforcement learning model. Moreover, ${\rm{EE}}_{l}^{^{(t)}} = \sum_{k \in \mathcal{K}}{\xi _{lk}} \rm{EE}$ represents the contribution value allocated by each agent in the total EE, where ${\xi _{lk}}$ is the contribution weight value, which is closely related to distance.

\subsubsection{Second-Layer: RIS Passive Beamforming Network}

Given the fixed AP selection configuration and multi-AP cooperative precoding, we focus on the RIS beamforming. In particular, the RIS-level controller is tasked with enhancing the transmission performance of the RIS cascaded link. The design of RIS-level control is based on the following steps.

\textbf{RIS-level agent}: Due to the assumption that the RIS is connected to the CPU and the phase shift of the RIS is directly controlled by it, we adopt the following strategy: We still assume that each AP is an agent, optimize the phase shift of the RIS in the second-layer network, and send the optimization result back to the CPU for selection. The CPU then selects the RIS phase that maximizes the reward in the current time slot.

\textbf{RIS-level observation space}:  States are defined as the overall system states. Here, we consider that the RIS is directly controlled by the CPU. Hence, the global CSI of the RIS-UE link and AP-RIS link is obtained and shared from the CPU for AP $l$ in time slot $t$, and we consider partial observation for AP $l$ in slot $t$ as
\begin{align}
    \mathcal{O}_{\mathrm{RIS}, l}^{^{(t)}} \!=\! \left(\mathbf{A}_{l}^{(t)},\mathbf{h}_{lk,d}^{(t)}, \mathbf{h}_{rk}^{(t)}, \mathbf{\Theta}^{(t)},\mathbf{G}_{lr}^{(t)},\mathbf{w}_{l}^{(t)}, {\rm{EE}}_{l}^{(t)} \right).
\end{align}
Then, we can obtain the fuzzy RIS-level observation, which is donated as $\hat{\mathcal{S}}_{\mathrm{RIS}}^{(t)} \!=\! ( \hat{\mathcal{S}}_{\mathrm{RIS},1}^{(t)},\dots,\hat{\mathcal{S}}_{\mathrm{RIS},F}^{(t)} )$, with each $\mathcal{S}_{\mathrm{RIS},f}^{(t)}$ being a random selection from the observed state, and $n$ representing the total number of fuzzy agents. The corresponding membership function is the same as those in the AP-level network.

\textbf{RIS-level action space}: Action  $a \in \mathcal{A}_{\mathrm{RIS}}$  for the phase shift matrix optimization is a one-hot mapping from the fuzzy action $\hat{\mathcal{A}}_{\mathrm{RIS}}^{(t)} = ( \hat{\mathcal{A}}_{\mathrm{RIS},1}^{(t)}, \dots, \hat{\mathcal{A}}_{\mathrm{RIS},F}^{(t)} )$ to the specific action $\mathcal{A}_{\mathrm{RIS}}^{(t)}$. Each AP has $\text{M}\times \text{K}\times \text{N} $ actions to select, in which the action pairs of the RIS-level policy at slot $t$ is denoted as $\Xi^{(t)}$, i.e.,  $\Xi^{(t)}=\{\Theta^{(t)},\mathbf{w}^{(t)}\}$.

\textbf{RIS-level reward space}: In accordance with the predefined reward function, the corresponding reward $r_{\mathrm{RIS}}^{(t)}$ at slot $t$ is determined as
\begin{equation}
r_{\mathrm{RIS},l}^{^{(t)}} = \frac{1}{K} \sum_{k \in \mathcal{K}} \mathbb{I}[\sum_{l\in \mathcal{A}_{k}} {\rm{EE}}_{l}^{^{(t)}}].
\end{equation}
In the context of environmental interaction, the utilization of fuzzy agents is imperative as opposed to conventional entities. This necessitates the implementation of fuzzification to compute the fuzzy reward $\hat{r}_{\mathrm{RIS}}^{(t)}=( \hat{r}_{\mathrm{RIS},1}^{(t)},\dots,\hat{r}_{\mathrm{RIS},F}^{(t)} )$ within the framework of reinforcement learning. The fuzzy reward can be mathematically represented as $\hat{r}_{\mathrm{RIS},l}^{(t)}=\sum_{k=1}^{K} \bar{\Xi}_{\mathrm{RIS},l,k}^{(t)} \times r_{\mathrm{RIS},k}^{(t)}$.

With the architecture of the FL-MARL, each fuzzy agent calculates its own policy gradient of the local actor network according to the joint abstract observation and action. Then, the objective function for the $i$-th policy $\pi_{\mathrm{AP},i}$ and $\pi_{\mathrm{RIS},i}$ can be designed as $L(\pi) \!=\! \sum_{\hat{s}_{i}^{(t)}}p_{\pi}(\hat{s}_{i}^{(t)})\sum_{\hat{a}_{i}^{(t)}}\pi(\hat{a}|\hat{s}_{i}^{(t)})\hat{r}_{i}^{(t)}$. Considering that the FL completes the mapping from the original entity to the fuzzy agent, the number of fuzzy agents can be determined based on network requirements to significantly reduce signaling overhead in the network and meet system deployment requirements.

Correspondingly, the $f$-th fuzzy reward $\hat{r}_{f}^{(t)}$ is based on the fuzzy action $\hat{\mathcal{A}}^{(t)}$ and state $\hat{\mathcal{S}}^{(t)}$ at both the AP-level and RIS-level, leading to the action values $Q{\pi_{\mathrm{AP}}}(\hat{s}_{\mathrm{AP}}^{(t)}, \hat{a}_{\mathrm{AP}}^{(t)})$ and $Q_{\pi_{\mathrm{RIS}}}(\hat{s}_{\mathrm{RIS}}^{(t)}, \hat{a}_{\mathrm{RIS}}^{(t)})$, respectively, which are calculated by the $i$-th critic network. The policy gradient of the local actor network for $\pi_{\mathrm{AP},i}$ and $\pi_{\mathrm{RIS},i}$ is denoted as
\begin{align}\label{Delta}
    \Delta_{\theta_{\pi_{i}}} \!\!J\left(\theta_{\pi_{i}}\right)\!\!=\!\!\sum_{\hat{a}_{i}^{(t)}} \!Q_{\pi}\!\!\left(\hat{s}^{(t)}, \hat{a}^{(t)}\!\right) \Delta_{\theta_{\pi_{i}}} \!\!\pi_{i}\left(\hat{a}_{i}^{(t)} \!\mid\! \hat{s}_{i}^{(t)} ; \theta_{\pi_{i}}\right) .
\end{align}
The $\Delta_{\theta_{\pi_{i}}} \pi_{i}(\hat{a}_{i}^{(t)} \mid \hat{s}_{i}^{(t)}; \theta_{\pi_{i}})$  in \eqref{Delta} denotes the output by the local and global policy network. Hence, the mean-squared Bellman error function for the joint critic network of the $f$-th fuzzy agent is denoted as
\begin{align}
L\left(\theta_{Q_{\pi}}\right)=\mathbb{E}\left[\left(Q_{\pi}\left(\hat{s}^{(t)}, \hat{a}^{(t)}\right)-y_{i}^{(t)}\right)^{2}\right],
\end{align}
with the local target  $y_{i}^{(t)}=\hat{r}_{i}^{(t)}+\gamma Q_{\pi}(\hat{s}^{(t+1)},\left.\hat{a}^{(t+1)}\right|{\hat{a}^{(t+1)} \sim \pi(\hat{s}^{(t+1)})})$.

The soft update is carried out in combination with the current network. The target actor network is $ \theta_{\pi_{i}^{\prime}} \leftarrow \tau \theta_{\pi_{i}^{\prime}}+(1-\tau) \theta_{\pi_{i}}$  and the target critic network is $ \theta_{Q_{\pi^{\prime}}} \leftarrow \tau \theta_{Q_{\pi^{\prime}}}+(1-\tau) \theta_{Q_{\pi}} $. Both the procedure of the AP-level and the RIS-level network of FL-MARL for maximizing the EE performance of the considered system are summarized in Algorithm 2.

Overall, we decouple the original problem into two sub-problems: AP selection and precoding, and RIS phase shift design, using a double-layer MADDPG network to address these issues. Specifically, dynamic AP selection based on local CSI and user location is implemented, and, based on AP selection, efficient joint precoding is designed to maximize the system's EE. Furthermore, in the context of an RIS-aided CF system with a large number of APs, the introduction of FL reduces the number of intelligent agents required in actual network calculations, thereby decreasing the network's computational complexity and providing practical guidance for engineering implementation.

\begin{table*}[t]
\caption{Comparison of Computational Complexity.}
	\label{tab2}
	\centering
	\footnotesize
	\renewcommand{\arraystretch}{2}
\begin{tabular}{lll}
    \hline\hline
    \textbf{\makecell{Parameters}} &\textbf{\makecell{1st Computational Complexity}} &\textbf{\makecell{2nd Computational Complexity}} \\ \hline\hline
    \multicolumn{2}{l}{\emph{\textbf{AP Selection with MADDPG}}} \\ \hline\hline
    \emph{\makecell{APS-MADDPG}}  & \makecell{$\mathcal{O}(L^2MKN^2\sum_{a=1}^{A_L}Q_a^{2}\!\!+\!\!L^2MK\sum_{c=1}^{C_L}Q_c^{2})$} & \makecell{$\mathcal{O}(L^2MKN^2\sum_{a=1}^{A_H}Q_a^{2}\!\!+\!\!LN\sum_{c=1}^{C_H}Q_c^{2} \!\!+\!\! L^2N_{AP}Q_{SE})$} \\ \hline
    \emph{\makecell{APS-FL-MADDPG}}  &\makecell{$\mathcal{O}(LN_FMKN^2\sum_{a=1}^{A_L}Q_a^{2}\!\!+\!\!LN_FMK\sum_{c=1}^{C_L}Q_c^{2})$} & \makecell{$\mathcal{O}(LN_FMKN^2\sum_{a=1}^{A_H}Q_a^{2}\!\!+\!\!LN\sum_{c=1}^{C_H}Q_c^{2} \!\!+\!\! LN_FN_{AP}Q_{SE})$} \\ \hline
    
    \multicolumn{2}{l}{\emph{\textbf{AP Full Coverage with MADDPG}}} \\ \hline\hline
    \emph{\makecell{APFC-MADDPG}} & \makecell{$\mathcal{O}(L^2MKN^2\sum_{a=1}^{A_L}Q_a^{2}\!\!+\!\!L^2MK\sum_{c=1}^{C_L}Q_c^{2})$} & \makecell{$\mathcal{O}(L^2MKN^2\sum_{a=1}^{A_H}Q_a^{2}\!\!+\!\!LN\sum_{c=1}^{C_H}Q_c^{2} \!\!+\!\! L^3Q_{SE})$}  \\ \hline
    \emph{\makecell{APFC-FL-MADDPG}} &\makecell{$\mathcal{O}(LN_FMKN^2\sum_{a=1}^{A_L}Q_a^{2}\!\!+\!\!LN_FMK\sum_{c=1}^{C_L}Q_c^{2})$} & \makecell{$\mathcal{O}(LN_FMKN^2\sum_{a=1}^{A_H}Q_a^{2}\!\!+\!\!LN\sum_{c=1}^{C_H}Q_c^{2} \!\!+\!\! L^2N_FQ_{SE})$} \\ \hline\hline
\end{tabular}
\end{table*}

\subsection{Comparative Analysis of Computational Complexity}
We compare the computational complexity of different positioning algorithms, as shown in Table \uppercase\expandafter{\romannumeral1}. For convenience of comparison, the following schemes are listed.
\subsubsection{\textbf{ZF}}
The RIS-aided CF mMIMO system adopts zero-forcing precoding with AP full coverage of all UEs.
\subsubsection{\textbf{AP Selection with MADDPG (APS-MADDPG)}}
The RIS-aided CF mMIMO system considers AP selection and joint precoding design with the MADDPG algorithm.
\subsubsection{\textbf{AP Full Coverage with MADDPG (APFC-MADDPG)}}
The RIS-aided CF mMIMO system considers AP full coverage of all UEs and joint precoding design with the MADDPG algorithm.
\subsubsection{\textbf{AP Selection with Fuzzy
Logic-based MADDPG (APS-FL-MADDPG)}}
The RIS-aided CF mMIMO system considers AP selection and joint precoding design with the FL-based MADDPG algorithm.
\subsubsection{\textbf{AP Full Coverage with Fuzzy
Logic-based MADDPG (APFC-FL-MADDPG)}}
The RIS-aided CF mMIMO system considers AP full coverage of all UEs and joint precoding design with the FL-based MADDPG algorithm.

For the various MADDPG-based algorithms proposed, the computational complexity is composed of actor network and critic network, respectively, where the input and output of the former are the state dimension $d_s$ and the action dimension $d_a$, while the input and output of the latter are the sum of the state and action dimensions $d_s + d_a$ and the critic value $d_c = 1$. Moreover, compared with the conventional MADDPG, the fuzzy logic technology used can reduce the input of the original actor network and critic network from the actual number of agents $L$ to the selected number of fuzzy agents $N_F$, thereby significantly reducing the computational complexity of the network to meet the needs of actual deployment as much as possible.

Therefore, the computational complexity of the first and second layer networks in APS-MADDPG can be denoted as $\mathcal{O}(L^2MKN^2\sum_{a=1}^{A_L}Q_a^{2}+L^2MK\sum_{c=1}^{C_L}Q_c^{2})$ and $\mathcal{O}(L^2MKN^2\sum_{a=1}^{A_H}Q_a^{2}+LN\sum_{c=1}^{C_H}Q_c^{2} + L^2N_{AP}Q_{SE})$, respectively, where $A_H$, $A_L$ and $C_H$, $C_L$ are the number of hidden layers for implementing the actor and critic networks. Also, $Q_a$ and $Q_c$ are the output size of the $a$-th and $c$-th layer or the input size of the next layer, respectively. Besides, $N_{AP}$ is the number of selected APs and $Q_{SE}$ represents the computational complexity of calculating the SE.
Compared with APS-MADDPG, the computational complexity of the APS-FL-MADDPG scheme is mainly determined by the number of fuzzy agents $N_{F}$, which can be represented as $\mathcal{O}(LN_FMKN^2\sum_{a=1}^{A_L}Q_a^{2}+LN_FMK\sum_{c=1}^{C_L}Q_c^{2})$ and $\mathcal{O}(LN_FMKN^2\sum_{a=1}^{A_H}Q_a^{2}+LN\sum_{c=1}^{C_H}Q_c^{2} + LN_FN_{AP}Q_{SE})$. Moreover, for the AP full coverage with MADDPG, the biggest difference in computational complexity compared to AP selection with MADDPG is that in the second layer network, the calculation of SE expression is updated from the original $N_{AP}$ to $L$. The specific comparison of the computational complexity of different algorithms is shown in Table \ref{tab2}.

\begin{rem}
Note that as observed from Table~\ref{tab2}, the computational complexity of MADDPG exhibits an exponential increase with the number of AP $L$ increasing. In contrast, upon integrating FL, the computational complexity increases linearly with the number of AP $L$. Correspondingly, the introduction of FL makes the computational complexity of FL-MADDPG linearly related to the number of fuzzy agents $N_{F}$.
\end{rem}

\begin{table}[t]
\centering
    \fontsize{9}{12}\selectfont
    \caption{The Model Structure and Experimental Details.}
    \label{Table1}
    \begin{tabular}{ccc}
    \toprule
    \bf Parameters &  \bf Size \\
    \midrule
    1st hidden layer & 256, Leaky Relu (0.01)\\
    2nd hidden layer & 128, Leaky Relu (0.01) \\
    Discounted factor $\gamma^{\mathrm{p}}$ and $\gamma^{\mathrm{f}}$ & 0.99 \\
    Experience pool size $\mathcal{D}^{\mathrm{p}}$ and $\mathcal{D}^{\mathrm{f}}$ & 512 and 512 \\
    Soft update rate $\tau^{\mathrm{p}}$ and $\tau^{\mathrm{f}}$ & 0.0003 and 0.02 \\
    \bottomrule
    \end{tabular}
\end{table}

\section{Numerical Simulation}
\subsection{Simulation Setup}
In this section, we evaluate the system performance of the joint AP selection and precoding scheme with different parameters. We consider an RIS-aided CF mMIMO system where all APs and UEs are uniformly distributed in a square area of size $100 \times 100$ $\text{m}^2$. We assume that there are $L=8$ APs and $K=6$ UEs. To enhance network capacity, the RIS is strategically deployed in the regional center. We adopt the channel model and path loss model similar to \cite{zhang2021joint}. We assume that the maximum transmit power for each AP is ${P_{\text{AP, max}}} = 5\, {\text{dB}}$. The transmission bandwidth is ${{B}} = 180\, \rm{kHz}$ Moreover, the model structure and experimental details of the first and second layers are shown in Table \ref{Table1}, and the common simulation parameters based on several existing works are listed in Table \ref{Table2}.

\subsection{Convergence of the Proposed Algorithm}
\begin{figure}[t]
\centering
\includegraphics[scale=0.5]{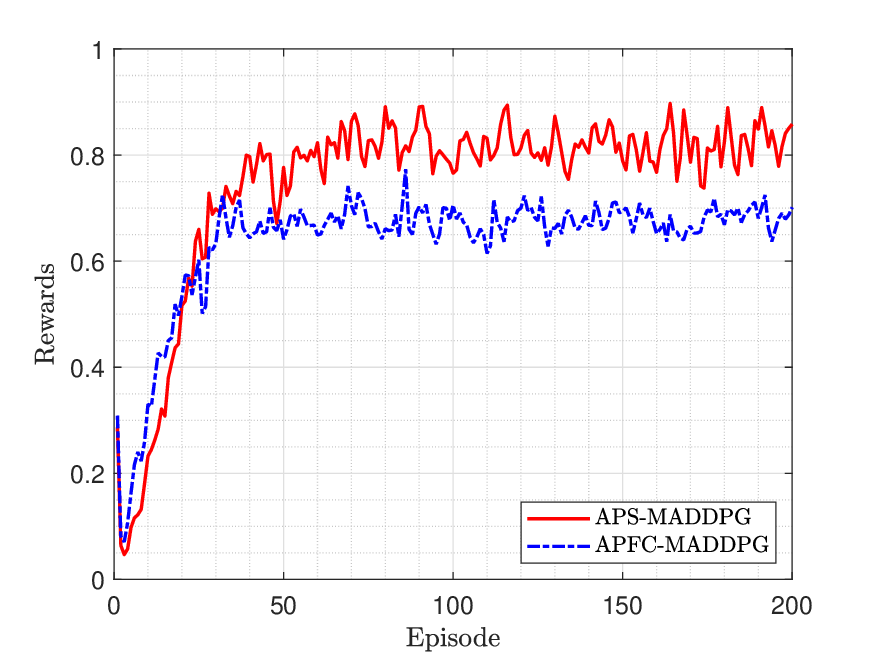}\vspace{-0.2cm}
\caption{Convergence rate of MADDGP for AP selection or AP full coverage scenarios ($L = 8$, $M = 2$, $K = 6$, $N = 64$, $P_{\rm{AP,max}} = 5\,\rm{dB}$, $P_{\rm{element}} = -20 \,\,\rm{dB}$).}\label{Fig_reward_MA}\vspace{-0.4cm}
\end{figure}

\begin{figure}[t]
\centering
\includegraphics[scale=0.5]{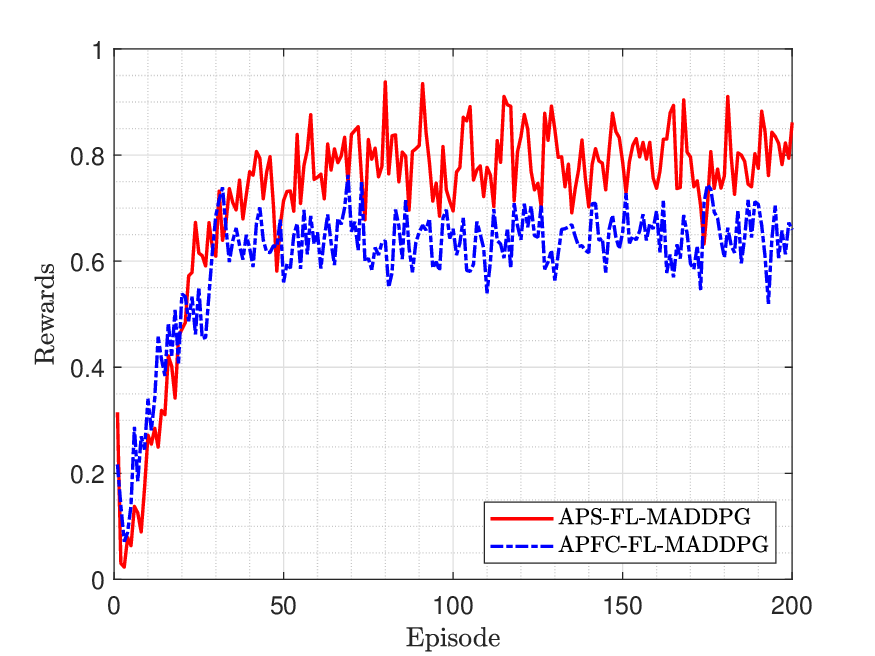}\vspace{-0.2cm}
\caption{Convergence rate of FL-based MADDGP for AP selection or AP full coverage scenarios ($L = 8$, $M = 2$, $K = 6$, $N = 64$, $P_{\rm{AP,max}} = 5\,\rm{dB}$, $P_{\rm{element}} = -20\,\,\rm{dB}$).}\label{Fig_reward_FL}\vspace{-0.4cm}
\end{figure}

In this subsection, we investigate the convergence rate over various algorithms with different AP selection scenarios in RIS-aided CF mMIMO systems. The convergence rate of MADDP and FL-MADDPG for AP selection and AP full coverage scenarios are shown in Fig.~\ref{Fig_reward_MA} and Fig.~\ref{Fig_reward_FL}. Specifically, we observe that the algorithms have all achieved convergence, with the number of times FL-MADDPG and MADDPG reach convergence being close. Also, it is noticeable that the FL-MADDPG algorithm exhibits greater fluctuations after convergence, which is due to the fuzzy uncertainty in the input-output mapping relationships introduced by FL-MADDPG. Meanwhile, we observe that the algorithm rewards after introducing the AP selection are higher than those with AP full coverage. This reveals that compared with the APFC-MADDPG, the proposed APS-MADDPG algorithm can further enhance the EE of the RIS-aided CF mMIMO system.

\begin{figure}[t]
\centering
\includegraphics[scale=0.5]{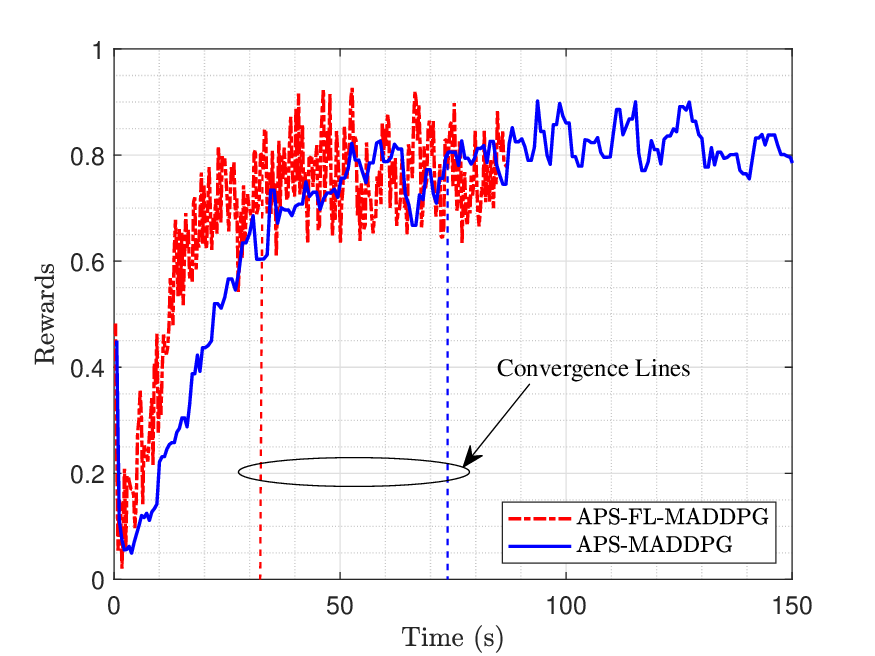}\vspace{-0.2cm}
\caption{Convergence time of MADDPG and FL-MADDPG algorithm with AP selection ($L = 8$, $M = 2$, $K = 6$, $N = 64$, $P_{\rm{AP,max}} = 5\,\,\rm{dB}$, $P_{\rm{element}} = -20\,\,\rm{dB}$).}\label{Fig_time}\vspace{-0.4cm}
\end{figure}

Fig.~\ref{Fig_time} shows a comparison of the convergence time of MADDPG and FL-MADDPG algorithms under the AP selection scenario. It is clear that the time required to achieve convergence of APS-FL-MADDPG is only 45\% of the time needed for APS-MADDPG to converge. As mentioned in the complexity analysis, this is because the introduction of FL significantly reduces the algorithmic complexity. However, the APS-FL-MADDPG algorithm results in some loss of channel state information, which leads to a decrease in system performance compared to APS-MADDPG. Therefore, when high precision is not a critical requirement for communication, we can significantly reduce the computational complexity and enhance computation speed at the cost of a minor performance loss by incorporating the FL algorithm.

\begin{table}[t]
\centering
    \fontsize{8}{12}\selectfont
    \caption{Simulation Parameters.}
    \label{Table2}
    \begin{tabular}{ccc}
    \toprule
    \bf Parameters &  \bf Values \\
    \midrule
    Noise power $\sigma^2$ & $-96\, \text{dBm}$\\
    Static power consumption of CPU $P^{\text{CPU}}$ & $5\,\text{W}$ \\
    Static power consumption of $l$-th AP $P_l^{\text{static}}$   & $50\,\text{mW}$ \\
    Static power consumption of $k$-th UE $P_k$   & $10\,\text{mW}$ \\
    Fixed power consumption of $l$-th fronthaul link  $P_{0,l}$ & $100\,\text{mW}$ \\
    Traffic-dependent power of $l$-th fronthaul link  $P_{{\rm{bt}},l}$ & \!\!\!\!$0.25\,\text{W}\!/\!(\rm{Gbits/s})$ \\
    Static power consumption of RIS element $P_{\text{element}}$ & $10\,\text{mW}$ \\
    Maximum transmit power of AP $P_{\text{AP,max}}$ & $3.2\,\text{W}$ \\
    \bottomrule
    \end{tabular}
\end{table}

\subsection{Impact of AP Transmission Power}
\begin{figure}[t]
\centering
\includegraphics[scale=0.5]{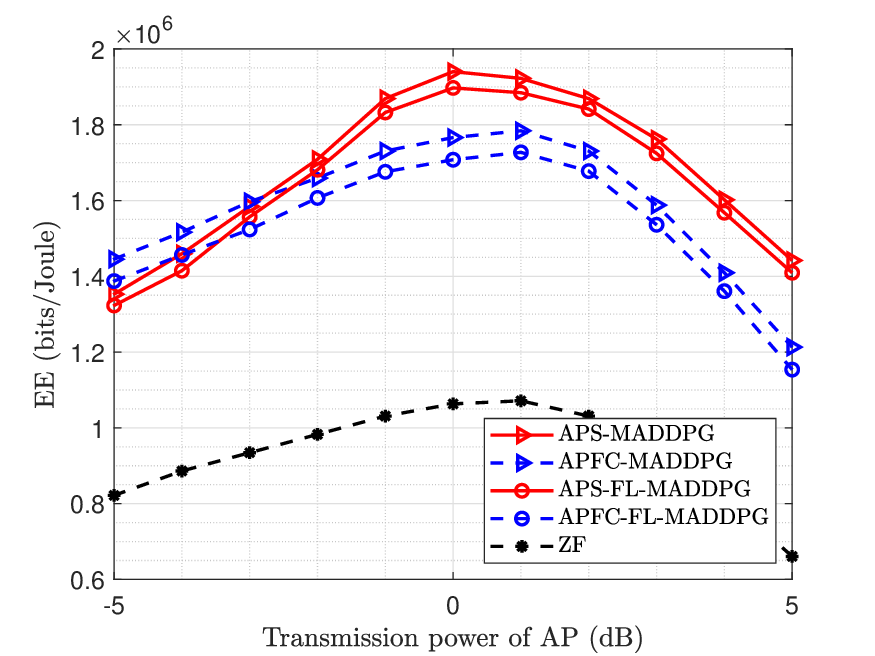}\vspace{-0.2cm}
\caption{Energy efficiency versus the transmission power constraints of APs with various algorithms ($L = 8$, $M = 2$, $K = 6$, $N = 64$, $P_{\rm{element}} = -20\,\,\rm{dB}$).}\label{Fig_power_EE}\vspace{-0.4cm}
\end{figure}

Fig.~\ref{Fig_power_EE} shows the EE of the RIS-aided CF mMIMO system versus the transmission power of APs with various algorithms. The results show that the EE initially increases and then decreases with the increase in transmission power, reaching its peak around the AP transmission power of $0\,\text{dB}$. At this point, the APS scheme improves the EE performance by 10\% compared to the APFC scheme. It is because when the AP transmission power is low, increasing it can significantly enhance performance. However, when the transmission power is sufficiently high, further increasing it would only result in a relatively small performance improvement, while the energy consumption increases substantially. Besides, we observe that at lower transmission powers, the EE performance of APFC surpasses that of the APS algorithm. This is because when the transmission power is low, AP full coverage is unable to meet the communication performance requirements of UEs. Under such circumstances, introducing AP selection results in even less fulfillment of user communication needs.

\begin{figure}[t]
\centering
\includegraphics[scale=0.5]{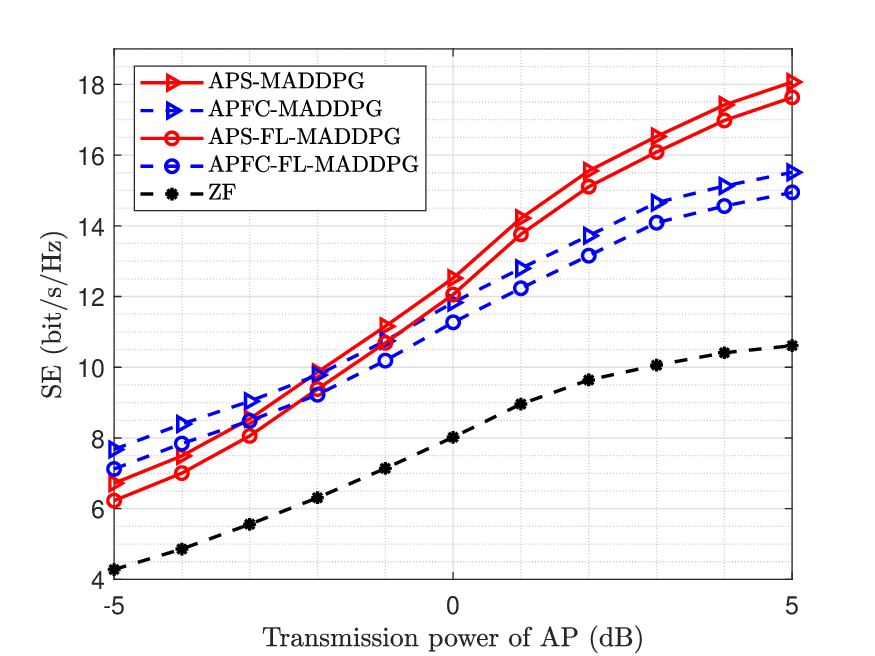}\vspace{-0.2cm}
\caption{Spectral efficiency versus the transmission power constraints of APs with various algorithms ($L = 8$, $M = 2$, $K = 6$, $N = 64$, $P_{\rm{element}} = -20\,\,\rm{dB}$).}\label{Fig_power_SE}\vspace{-0.4cm}
\end{figure}

Fig.~\ref{Fig_power_SE} shows the SE of the RIS-aided CF mMIMO system versus the transmission power of APs with various algorithms. We observe that the SE performance of the proposed MADDPG algorithm far exceeds that of the ZF algorithm. Besides, the results show that increasing the AP transmission power is always beneficial for SE, but with diminishing returns when the AP transmission power is sufficiently large. Also, it is clear that as the AP transmission power increases, the advantages of the APS algorithm become more prominent compared with the APFC algorithm. Based on the above findings, it is essential to design the transmission power of the transceivers reasonably according to the different requirements of SE/EE. Additionally, the significance of AP selection becomes more apparent in high-power transmission scenarios.

\subsection{Impact of the Number of AP Antennas}
\begin{figure}[t]
\centering
\includegraphics[scale=0.5]{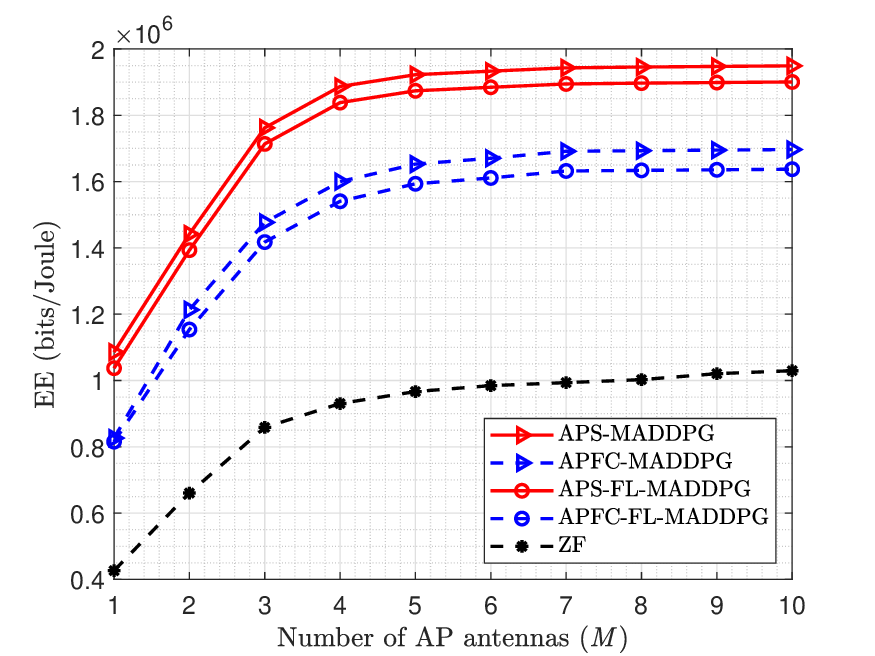}\vspace{-0.2cm}
\caption{Energy efficiency versus the number of AP antennas with various algorithms ($L = 8$, $K = 6$, $N = 64$, $P_{\rm{AP,max}} = 5\,\,\rm{dB}$, $P_{\rm{element}} = -20\,\,\rm{dB}$).}\label{Fig_antenna_EE}\vspace{-0.4cm}
\end{figure}

Fig.~\ref{Fig_antenna_EE} shows the EE versus the number of AP antennas with various algorithms. It is clear that the EE increases with the number of AP antennas increasing. However, when the number of AP antennas exceeds 6, the performance gains from further increasing the number of AP antennas are marginal. When the number of AP antennas is $M = 10$, the EE performance of the APS-FL-MADDPG algorithm improves by 16\% and 85\% compared to the EE performance of APFC-FL-MADDPG and ZF algorithms, respectively. Meanwhile, compared with the MADDPG algorithm, the performance of the FL-MADDPG algorithm under AP selection or AP full coverage scenarios only decreases by 2.5\% and 3.6\%, respectively. 

\begin{figure}[t]
\centering
\includegraphics[scale=0.5]{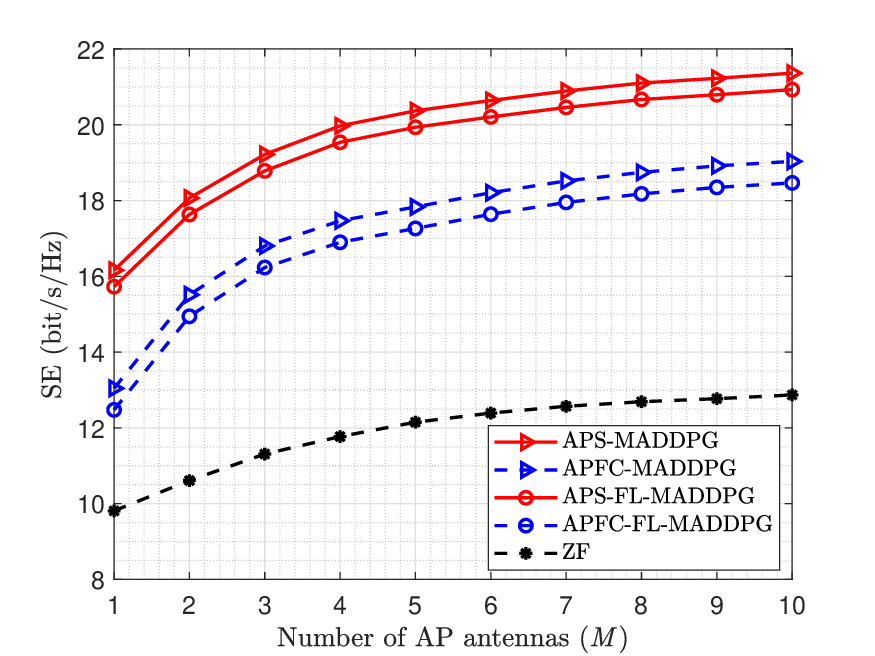}\vspace{-0.2cm}
\caption{Spectral efficiency versus the number of AP antennas with various algorithms ($L = 8$, $K = 6$, $N = 64$, $P_{\rm{AP,max}} = 5\,\,\rm{dB}$, $P_{\rm{element}} = -20\,\,\rm{dB}$).}\label{Fig_antenna_SE}\vspace{-0.4cm}
\end{figure}

Fig.~\ref{Fig_antenna_SE} shows the SE versus the number of AP antennas with various algorithms. We can observe that the SE of the RIS-aided CF mMIMO system increases with the number of AP antennas increasing, but with diminishing returns when the number of AP antennas is sufficiently large which can serve as supporting evidence for the trend in EE as shown in Fig.~\ref{Fig_antenna_EE}. Also, the APS-MADDPG algorithm outperforms the APFC-MADDPG algorithm by 23\% and 12\% in SE performance when $M = 1$ and $M = 10$, respectively. This reveals that in the RIS-aided CF mMIMO system when the number of AP antennas is limited, considering AP selection to enhance system performance is a viable option.

\subsection{Impact of the Number of RIS Elements}
\begin{figure}[t]
\centering
\includegraphics[scale=0.5]{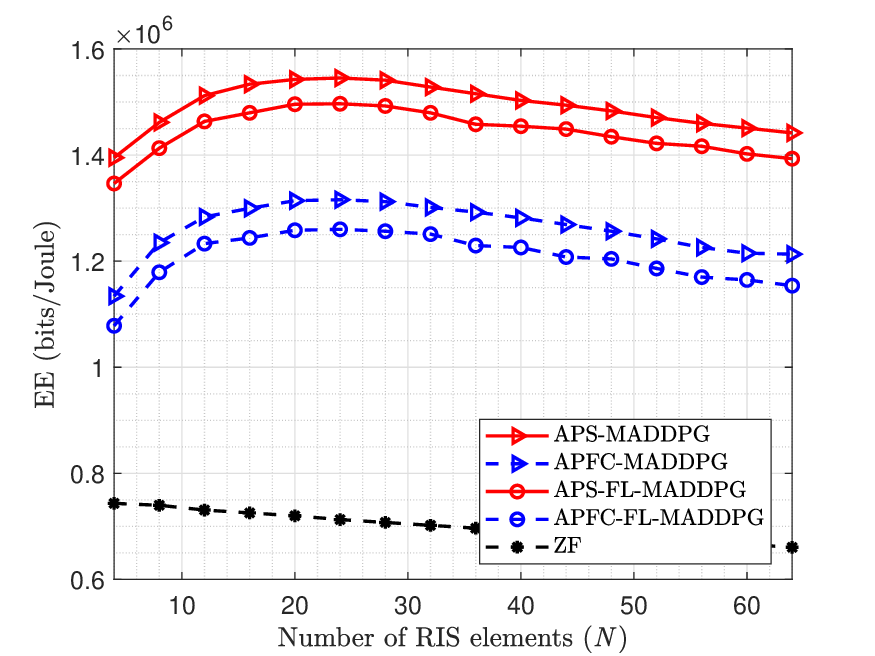}\vspace{-0.2cm}
\caption{Energy efficiency versus the number of RIS elements with various algorithms ($L = 8$, $M = 2$, $K = 6$, $P_{\rm{AP,max}} = 5\,\,\rm{dB}$, $P_{\rm{element}} = -20\,\,\rm{dB}$).}\label{Fig_element_EE}\vspace{-0.4cm}
\end{figure}

Fig.~\ref{Fig_element_EE} shows the EE versus the number of RIS elements with various algorithms. It is clear that the EE of the RIS-aided CF mMIMO system with MADDPG initially increases and then decreases with the increase in the number of RIS elements, reaching its peak at around $N = 24$. However, the EE with ZF continues to decrease as the number of RIS elements increases. Also, the EE performance under the AP selection algorithm consistently outperforms the AP full coverage algorithm, and the FL algorithm exhibits minimal performance loss. Besides, the APS-MADDPG algorithm outperforms the APFC-MADDPG algorithm by 23\%, 17\%, and 19\% in SE performance when $N = 4$, $N = 24$, and $N = 64$, respectively. It reveals that in the RIS-aided CF mMIMO system when the number of RIS elements is limited, considering AP selection to enhance system performance is a viable option. 

\begin{figure}[t]
\centering
\includegraphics[scale=0.5]{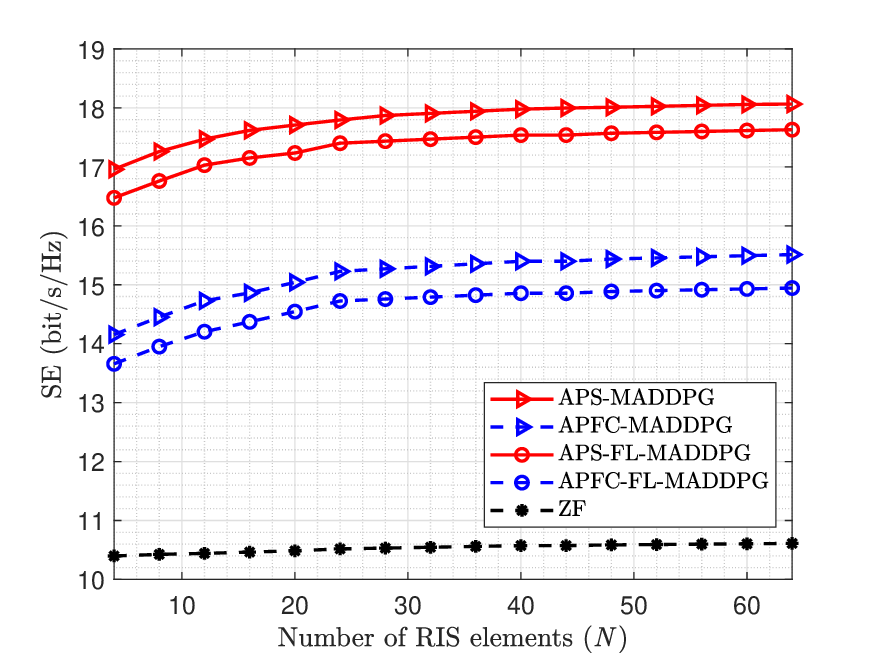}\vspace{-0.2cm}
\caption{Spectral efficiency versus the number of RIS elements with various algorithms ($L = 8$, $M = 2$, $K = 6$, $P_{\rm{AP,max}} = 5\,\,\rm{dB}$, $P_{\rm{element}} = -20\,\,\rm{dB}$).}\label{Fig_element_SE}\vspace{-0.4cm}
\end{figure}

Fig.~\ref{Fig_element_SE} shows the SE versus the number of RIS elements with various algorithms. We can observe that the SE of the RIS-aided CF mMIMO system with MADDPG increases with the number of RIS elements increased. However, the improvement of EE performance with ZF is marginal, which is the reason for the decrease in EE with the ZF algorithm, as shown in Fig.~\ref{Fig_element_EE}. Besides, as the number of RIS elements increases from $N = 24$ to $N = 64$, SE based on the APS-MADDPG algorithm and SE based on the APFC-MADDPG algorithm grow by 17\% and 16\%, respectively. This indicates that when we aim to maximize EE with different numbers of RIS elements, the SE gap change is not significant regardless of the AP selection.

\begin{figure}[t]
\centering
\includegraphics[scale=0.5]{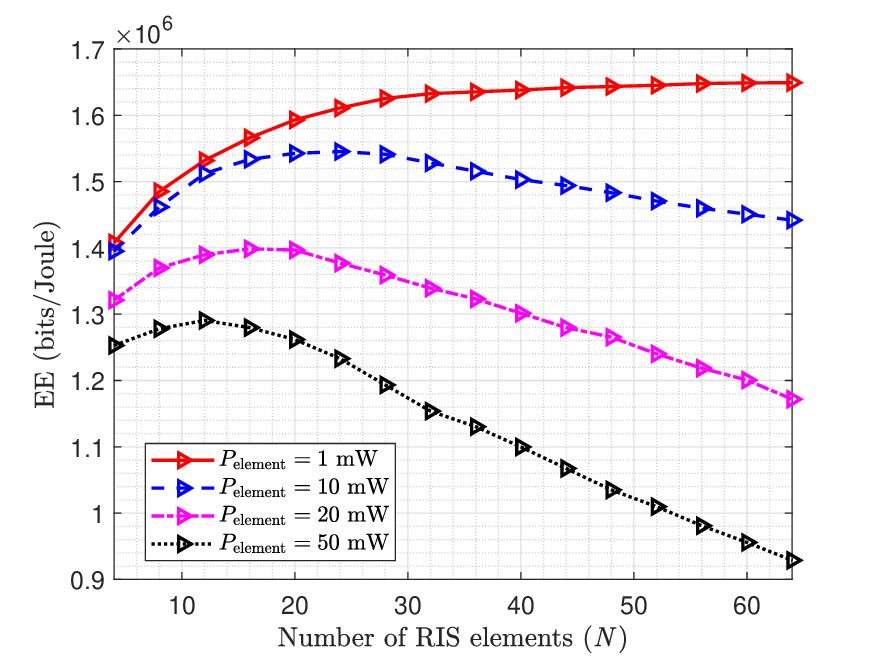}\vspace{-0.2cm}
\caption{Energy efficiency versus different static power consumption of RIS element with APS-MADDPG algorithm ($L = 8$, $M = 2$, $K = 6$, $P_{\rm{AP,max}} = 5\,\rm{dB}$).}\label{Fig_elementpower}\vspace{-0.4cm}
\end{figure}

Fig.~\ref{Fig_elementpower} shows the EE versus different static power consumption of the RIS elements with the APS-MADDPG algorithm. It is clear that when adopting the APS-MADDPG algorithm, the system EE performance initially increases and then decreases with the increase in the number of RIS elements. When the RIS elements power is at ${P_{\rm{element}}}=10,\, 20,\, 50\, \rm{mW}$, the EE peak values are at ${N = 24,\,16,\,12}$, respectively. It is important to note that with the increase in the power consumption of the RIS elements, the EE turning point corresponds to a smaller number of RIS elements. Therefore, in RIS-aided CF mMIMO systems, it is essential to design the power consumption of RIS elements thoughtfully to achieve optimal EE performance.

\section{Conclusion}
In this paper, we focused on a user-centric RIS-aided CF mMIMO system and investigated the joint precoding and AP selection scenarios. Then, we proposed a novel double-layer MARL scheme to address the multivariate nonconvex maximization EE problem. Moreover, an adaptive power threshold-based AP selection scheme was proposed to further reduce the energy consumption of the considered system. To reduce the computational complexity and accelerate convergence, a FL strategy was introduced to the double-layer MARL framework. The results showed that the proposed AP selection algorithm can enhance system performance by 10\%, and the improvement is more significant as the transmission power of the AP increases. Furthermore, the proposed FL-based MARL scheme can significantly reduce the computational complexity while closely approximating the performance of traditional MARL, thereby attaining an excellent balance between convergence speed and EE performance of RIS-aided CF mMIMO systems.

\bibliographystyle{IEEEtran}
\bibliography{IEEEabrv,Ref}

\begin{thebibliography}{10}
\providecommand{\url}[1]{#1}
\csname url@samestyle\endcsname
\providecommand{\newblock}{\relax}
\providecommand{\bibinfo}[2]{#2}
\providecommand{\BIBentrySTDinterwordspacing}{\spaceskip=0pt\relax}
\providecommand{\BIBentryALTinterwordstretchfactor}{4}
\providecommand{\BIBentryALTinterwordspacing}{\spaceskip=\fontdimen2\font plus
\BIBentryALTinterwordstretchfactor\fontdimen3\font minus
  \fontdimen4\font\relax}
\providecommand{\BIBforeignlanguage}[2]{{%
\expandafter\ifx\csname l@#1\endcsname\relax
\typeout{** WARNING: IEEEtran.bst: No hyphenation pattern has been}%
\typeout{** loaded for the language `#1'. Using the pattern for}%
\typeout{** the default language instead.}%
\else
\language=\csname l@#1\endcsname
\fi
#2}}
\providecommand{\BIBdecl}{\relax}
\BIBdecl

\bibitem{wang2023road}
C.-X. Wang, X.~You, X.~Gao, X.~Zhu, Z.~Li, C.~Zhang, H.~Wang, Y.~Huang,
  Y.~Chen, H.~Haas \emph{et~al.}, ``On the road to {6G}: Visions, requirements,
  key technologies and testbeds,'' \emph{IEEE Commun. Surveys Tuts.}, vol.~25,
  no.~2, pp. 905--974, Feb. 2023.

\bibitem{you2024next}
C.~You, Y.~Cai, Y.~Liu, M.~Di~Renzo, T.~M. Duman, A.~Yener, and A.~L.
  Swindlehurst, ``Next generation advanced transceiver technologies for {6G},''
  \emph{arXiv preprint arXiv:2403.16458}, 2024.

\bibitem{larsson2014massive}
E.~G. Larsson, O.~Edfors, F.~Tufvesson, and T.~L. Marzetta, ``Massive {MIMO}
  for next generation wireless systems,'' \emph{IEEE Commun. Mag.}, vol.~52,
  no.~2, pp. 186--195, Feb. 2014.

\bibitem{wu2021comprehensive}
Q.~Wu, J.~Xu, Y.~Zeng, D.~W.~K. Ng, N.~Al-Dhahir, R.~Schober, and A.~L.
  Swindlehurst, ``A comprehensive overview on {5G}-and-beyond networks with
  {UAVs}: From communications to sensing and intelligence,'' \emph{IEEE J. Sel.
  Areas Commun.}, vol.~39, no.~10, pp. 2912--2945, Oct. 2021.

\bibitem{bjornson2016deploying}
E.~Bj{\"o}rnson, L.~Sanguinetti, and M.~Kountouris, ``Deploying dense networks
  for maximal energy efficiency: Small cells meet massive {MIMO},'' \emph{IEEE
  J. Sel. Areas Commun.}, vol.~34, no.~4, pp. 832--847, Apr. 2016.

\bibitem{buzzi2019user}
S.~Buzzi, C.~D¡¯Andrea, A.~Zappone, and C.~D¡¯Elia, ``User-centric {5G}
  cellular networks: Resource allocation and comparison with the cell-free
  massive {MIMO} approach,'' \emph{IEEE Trans. Wireless Commun.}, vol.~19,
  no.~2, pp. 1250--1264, Feb. 2019.

\bibitem{zhang2020prospective}
J.~Zhang, E.~Bj{\"o}rnson, M.~Matthaiou, D.~W.~K. Ng, H.~Yang, and D.~J. Love,
  ``Prospective multiple antenna technologies for beyond {5G},'' \emph{IEEE J.
  Sel. Areas Commun.}, vol.~38, no.~8, pp. 1637--1660, Aug. 2020.

\bibitem{ngo2017cell}
H.~Q. Ngo, A.~Ashikhmin, H.~Yang, E.~G. Larsson, and T.~L. Marzetta,
  ``Cell-free massive {MIMO} versus small cells,'' \emph{IEEE Trans. Wireless
  Commun.}, vol.~16, no.~3, pp. 1834--1850, Mar. 2017.

\bibitem{nayebi2017precoding}
E.~Nayebi, A.~Ashikhmin, T.~L. Marzetta, H.~Yang, and B.~D. Rao, ``Precoding
  and power optimization in cell-free massive {MIMO} systems,'' \emph{IEEE
  Trans. Wireless Commun.}, vol.~16, no.~7, pp. 4445--4459, Jul. 2017.

\bibitem{bashar2019energy}
M.~Bashar, K.~Cumanan, A.~G. Burr, H.~Q. Ngo, E.~G. Larsson, and P.~Xiao,
  ``Energy efficiency of the cell-free massive {MIMO} uplink with optimal
  uniform quantization,'' \emph{IEEE Transa. Green Commun. Networking}, vol.~3,
  no.~4, pp. 971--987, Apr. 2019.

\bibitem{van2021reconfigurable}
T.~Van~Chien, H.~Q. Ngo, S.~Chatzinotas, M.~Di~Renzo, and B.~Ottersten,
  ``Reconfigurable intelligent surface-assisted cell-free massive {MIMO}
  systems over spatially-correlated channels,'' \emph{IEEE Trans. Wireless
  Commun.}, vol.~21, no.~7, pp. 5106--5128, Jul. 2021.

\bibitem{verma2020toward}
S.~Verma, S.~Kaur, M.~A. Khan, and P.~S. Sehdev, ``Toward green communication
  in {6G}-enabled massive internet of things,'' \emph{IEEE Int. Things J.},
  vol.~8, no.~7, pp. 5408--5415, Jul. 2020.

\bibitem{9140329}
M.~Di~Renzo, A.~Zappone, M.~Debbah, M.-S. Alouini, C.~Yuen, J.~de~Rosny, and
  S.~Tretyakov, ``Smart radio environments empowered by reconfigurable
  intelligent surfaces: How it works, state of research, and the road ahead,''
  \emph{IEEE J. Sel. Areas Commun.}, vol.~38, no.~11, pp. 2450--2525, Nov.
  2020.

\bibitem{10555049}
Q.~Wu, B.~Zheng, C.~You, L.~Zhu, K.~Shen, X.~Shao, W.~Mei, B.~Di, H.~Zhang,
  E.~Basar, L.~Song, M.~D. Renzo, Z.-Q. Luo, and R.~Zhang, ``Intelligent
  surfaces empowered wireless network: Recent advances and the road to {6G},''
  \emph{Proc. IEEE, early access}, 2024.

\bibitem{pan2021reconfigurable}
C.~Pan, H.~Ren, K.~Wang, J.~F. Kolb, M.~Elkashlan, M.~Chen, M.~Di~Renzo,
  Y.~Hao, J.~Wang, A.~L. Swindlehurst \emph{et~al.}, ``Reconfigurable
  intelligent surfaces for {6G} systems: Principles, applications, and research
  directions,'' \emph{IEEE Commun. Mag.}, vol.~59, no.~6, pp. 14--20, Jun.
  2021.

\bibitem{zhu2024marl}
Y.~Zhu, E.~Shi, Z.~Liu, J.~Zhang, and B.~Ai, ``Multi-agent reinforcement
  learning-based joint precoding and phase shift optimization for {RIS}-aided
  cell-free massive {MIMO} systems,'' \emph{IEEE Trans. Veh. Technol., early
  access}, 2024.

\bibitem{cai2021intelligent}
Y.~Cai, M.-M. Zhao, K.~Xu, and R.~Zhang, ``Intelligent reflecting surface aided
  full-duplex communication: Passive beamforming and deployment design,''
  \emph{IEEE Trans. Wireless Commun.}, vol.~21, no.~1, pp. 383--397, Jan. 2021.

\bibitem{zuo2020resource}
J.~Zuo, Y.~Liu, Z.~Qin, and N.~Al-Dhahir, ``Resource allocation in intelligent
  reflecting surface assisted {NOMA} systems,'' \emph{IEEE Trans. Commun.},
  vol.~68, no.~11, pp. 7170--7183, Nov. 2020.

\bibitem{li2021robust}
S.~Li, B.~Duo, M.~Di~Renzo, M.~Tao, and X.~Yuan, ``Robust secure {UAV}
  communications with the aid of reconfigurable intelligent surfaces,''
  \emph{IEEE Trans. Wireless Commun.}, vol.~20, no.~10, pp. 6402--6417, Oct.
  2021.

\bibitem{10556753}
E.~Shi, J.~Zhang, H.~Du, B.~Ai, C.~Yuen, D.~Niyato, K.~B. Letaief, and X.~Shen,
  ``{RIS}-aided cell-free massive {MIMO} systems for {6G}: Fundamentals, system
  design, and applications,'' \emph{Proc. IEEE}, vol. 112, no.~4, pp. 331--364,
  Apr. 2024.

\bibitem{zhang2021joint}
Z.~Zhang and L.~Dai, ``A joint precoding framework for wideband reconfigurable
  intelligent surface-aided cell-free network,'' \emph{IEEE Trans. Signal
  Process.}, vol.~69, pp. 4085--4101, Jun. 2021.

\bibitem{ma2022cooperative}
X.~Ma, D.~Zhang, M.~Xiao, C.~Huang, and Z.~Chen, ``Cooperative beamforming for
  {RIS}-aided cell-free massive {MIMO} networks,'' \emph{IEEE Trans. Wireless
  Commun.}, vol.~22, no.~11, pp. 7243--7258, Nov. 2023.

\bibitem{ozccelikkale2015linear}
A.~{\"O}z{\c{c}}elikkale and T.~M. Duman, ``Linear precoder design for
  simultaneous information and energy transfer over two-user {MIMO}
  interference channels,'' \emph{IEEE Trans. Wireless Commun.}, vol.~14,
  no.~10, pp. 5836--5847, Oct. 2015.

\bibitem{ghiasi2022energy}
N.~Ghiasi, S.~Mashhadi, S.~Farahmand, S.~M. Razavizadeh, and I.~Lee, ``Energy
  efficient {AP} selection for cell-free massive {MIMO} systems: Deep
  reinforcement learning approach,'' \emph{IEEE Trans. Green Commun.
  Networking}, vol.~7, no.~1, pp. 29--41, 2022.

\bibitem{le2021energy}
Q.~N. Le, V.-D. Nguyen, O.~A. Dobre, and R.~Zhao, ``Energy efficiency
  maximization in {RIS}-aided cell-free network with limited backhaul,''
  \emph{IEEE Commun. Lett.}, vol.~25, no.~6, pp. 1974--1978, Jun. 2021.

\bibitem{lyu2023energy}
W.~Lyu, Y.~Xiu, S.~Yang, C.~Yuen, and Z.~Zhang, ``Energy-efficient cell-free
  network assisted by hybrid {RISs},'' \emph{IEEE Wireless Commun. Lett.},
  vol.~12, no.~4, pp. 718--722, Apr. 2023.

\bibitem{siddiqi2022energy}
M.~Z. Siddiqi, R.~Mackenzie, M.~Hao, and T.~Mir, ``On energy efficiency of
  wideband {RIS}-aided cell-free network,'' \emph{IEEE Access}, vol.~10, pp.
  19\,742--19\,752, 2022.

\bibitem{chuang2023deep}
Y.-C. Chuang, W.-Y. Chiu, R.~Y. Chang, and Y.-C. Lai, ``Deep reinforcement
  learning for energy efficiency maximization in cache-enabled cell-free
  massive {MIMO} networks: Single-and multi-agent approaches,'' \emph{IEEE
  Trans. Veh. Technol.}, vol.~78, no.~8, pp. 10\,826--10\,839, Aug. 2023.

\bibitem{huang2020reconfigurable}
C.~Huang, R.~Mo, and C.~Yuen, ``Reconfigurable intelligent surface assisted
  multiuser miso systems exploiting deep reinforcement learning,'' \emph{IEEE
  J. Sel. Areas Commun.}, vol.~38, no.~8, pp. 1839--1850, Aug. 2020.

\bibitem{chen2023distributed}
C.~Chen, S.~Xu, J.~Zhang, and J.~Zhang, ``A distributed machine learning-based
  approach for {IRS}-enhanced cell-free {MIMO} networks,'' \emph{IEEE Trans.
  Wireless Commun.}, Oct. 2023.

\bibitem{zhang2021multi}
K.~Zhang, Z.~Yang, and T.~Ba{\c{s}}ar, ``Multi-agent reinforcement learning: A
  selective overview of theories and algorithms,'' \emph{Handbook of
  reinforcement learning and control}, pp. 321--384, 2021.

\bibitem{rahmani2022multi}
M.~Rahmani, M.~J. Dehghani, P.~Xiao, M.~Bashar, and M.~Debbah, ``Multi-agent
  reinforcement learning-based pilot assignment for cell-free massive {MIMO}
  systems,'' \emph{IEEE Access}, vol.~10, pp. 120\,492--120\,502, 2022.

\bibitem{liu2023double}
Z.~Liu, J.~Zhang, Z.~Liu, H.~Xiao, and B.~Ai, ``Double-layer power control for
  mobile cell-free {XL-MIMO} with multi-agent reinforcement learning,''
  \emph{IEEE Trans. Wireless Commun.}, Oct. 2023.

\bibitem{banerjee2023access}
B.~Banerjee, R.~C. Elliott, W.~A. Krzymie{\'n}, and M.~Medra, ``Access point
  clustering in cell-free massive {MIMO} using conventional and federated
  multi-agent reinforcement learning,'' \emph{IEEE Trans. Mach. Learn. Commun.
  Networking}, vol.~1, no.~6, pp. 107--123, Jun. 2023.

\bibitem{liu2023uplink}
Z.~Liu, Z.~Liu, J.~Zhang, H.~Xiao, B.~Ai, and D.~W.~K. Ng, ``Uplink power
  control for extremely large-scale {MIMO} with multi-agent reinforcement
  learning and fuzzy logic,'' \emph{arXiv preprint arXiv:2302.09290}, 2023.

\bibitem{tilahun2023multi}
F.~D. Tilahun, A.~T. Abebe, and C.~G. Kang, ``Multi-agent reinforcement
  learning for distributed resource allocation in cell-free massive
  {MIMO}-enabled mobile edge computing network,'' \emph{IEEE Trans. Veh.
  Technol.}, Jun. 2023.

\bibitem{abdallah2024multi}
A.~Abdallah, A.~Celik, M.~M. Mansour, and A.~M. Eltawil, ``Multi-agent deep
  reinforcement learning for beam codebook design in {RIS}-aided systems,''
  \emph{IEEE Trans. Wireless Commun., early access}, 2024.

\bibitem{9733984}
L.~Qian, P.~Yang, M.~Xiao, O.~A. Dobre, M.~D. Renzo, J.~Li, Z.~Han, Q.~Yi, and
  J.~Zhao, ``Distributed learning for wireless communications: Methods,
  applications and challenges,'' \emph{IEEE J. Sel. Top. Signal Process.},
  vol.~16, no.~3, pp. 326--342, Mar. 2022.

\bibitem{yu2021smart}
X.~Yu, V.~Jamali, D.~Xu, D.~W.~K. Ng, and R.~Schober, ``Smart and
  reconfigurable wireless communications: From {IRS} modeling to algorithm
  design,'' \emph{IEEE Wireless Commun.}, vol.~28, no.~6, pp. 118--125, Jun.
  2021.

\bibitem{bjornson2019making}
E.~Bj{\"o}rnson and L.~Sanguinetti, ``{Making cell-free massive MIMO
  competitive with MMSE processing and centralized implementation},''
  \emph{IEEE Trans. Wireless Commun.}, vol.~19, no.~1, pp. 77--90, Jan. 2019.

\bibitem{bashar2021limited}
M.~Bashar, P.~Xiao, R.~Tafazolli, K.~Cumanan, A.~G. Burr, and E.~Bj{\"o}rnson,
  ``Limited-fronthaul cell-free massive {MIMO} with local mmse receiver under
  rician fading and phase shifts,'' \emph{IEEE Wireless Commun. Lett.},
  vol.~10, no.~9, pp. 1934--1938, Sep. 2021.

\bibitem{ngo2017total}
H.~Q. Ngo, L.-N. Tran, T.~Q. Duong, M.~Matthaiou, and E.~G. Larsson, ``On the
  total energy efficiency of cell-free massive {MIMO},'' \emph{IEEE Trans.
  Green Commun. Netw.}, vol.~2, no.~1, pp. 25--39, Jan. 2017.

\bibitem{huang2019reconfigurable}
C.~Huang, A.~Zappone, G.~C. Alexandropoulos, M.~Debbah, and C.~Yuen,
  ``Reconfigurable intelligent surfaces for energy efficiency in wireless
  communication,'' \emph{IEEE Trans. Wireless Commun.}, vol.~18, no.~8, pp.
  4157--4170, Aug. 2019.

\bibitem{van2020power}
T.~Van~Chien, T.~N. Canh, E.~Bj{\"o}rnson, and E.~G. Larsson, ``Power control
  in cellular massive {MIMO} with varying user activity: A deep learning
  solution,'' \emph{IEEE Trans. Wireless Commun.}, vol.~19, no.~9, pp.
  5732--5748, Sep. 2020.

\bibitem{chafii2023emergent}
M.~Chafii, S.~Naoumi, R.~Alami, E.~Almazrouei, M.~Bennis, and M.~Debbah,
  ``Emergent communication in multi-agent reinforcement learning for future
  wireless networks,'' \emph{IEEE Int. Things Mag.}, vol.~6, no.~4, pp. 18--24,
  Apr. 2023.

\bibitem{hajek2013metamathematics}
P.~H{\'a}jek, \emph{Metamathematics of fuzzy logic}.\hskip 1em plus 0.5em minus
  0.4em\relax Springer Science \& Business Media, 2013, vol.~4.

\end{thebibliography}

\end{document}